\documentclass[twocolumn, times, linenumbers]{aastex63}
\usepackage{hyperref}
\usepackage{physics}
\hypersetup{colorlinks, citecolor=blue, linkcolor=blue, urlcolor=blue}
\usepackage{comment}

\usepackage{CJK}

\received{}
\revised{}
\accepted{}
\submitjournal{AAS Journals}

\shorttitle{A Parameter Survey of CPDs}
\shortauthors{Sagynbayeva, Li, et al.}
\graphicspath{{./}{./Figures/}}

\begin{document}
\nolinenumbers
\begin{CJK*}{UTF8}{gbsn}

\title{Circumplanetary Disks are Rare around Planets at Large Orbital Radii:\\ A Parameter Survey of Flow Morphology around Giant Planets}

%\correspondingauthor{}
%\email{}

\author[0000-0002-6650-3829]{Sabina Sagynbayeva}
\affiliation{Department of Physics and Astronomy, Stony Brook University, Stony Brook, NY 11794, USA}

\author[0000-0001-9222-4367]{Rixin Li(李日新)}
\altaffiliation{51 Pegasi b Fellow}
\affiliation{Department of Astronomy, Theoretical Astrophysics Center, and Center for Integrative Planetary Science, University of California Berkeley, Berkeley, CA 94720-3411, USA}

\author[0000-0002-6946-6787]{Aleksandra Kuznetsova}
\affiliation{Center for Computational Astrophysics, Flatiron Institute, 162 Fifth Avenue, New York, NY 10010, USA}
\affiliation{American Museum of Natural History, 200 Central Park West, New York, New York, 10024, USA}

\author[0000-0003-3616-6822]{Zhaohuan Zhu(朱照寰)}
\affiliation{Department of Physics and Astronomy, University of Nevada, Las Vegas, 4505 S.~Maryland Parkway, Las Vegas, NV~89154-4002, USA}

\author[0000-0002-2624-3399]{Yan-Fei Jiang(姜燕飞)}
\affiliation{Center for Computational Astrophysics, Flatiron Institute, 162 Fifth Avenue, New York, NY 10010, USA}

\author[0000-0001-5032-1396]{Philip J. Armitage}
\affiliation{Department of Physics and Astronomy, Stony Brook University, Stony Brook, NY 11794, USA}
\affiliation{Center for Computational Astrophysics, Flatiron Institute, 162 Fifth Avenue, New York, NY 10010, USA}

\begin{abstract}
\nolinenumbers
We investigate how the formation and structure of circumplanetary disks (CPDs) varies with planet mass 
and protoplanetary disk aspect ratio. Using static mesh refinement and a near-isothermal equation of state, we perform a small parameter survey of hydrodynamic simulations with parameters appropriate for disk-embedded protoplanets at moderate to large orbital radii. We find that CPD formation occurs along a continuum, with ``diskiness'' increasing smoothly with planetary mass and decreasing disk aspect ratio. As expected from disk hydrostatic equilibrium arguments, the transition from envelope-dominated to disk-dominated structures is determined to first order by the ratio of the planetary Hill sphere radius to the disk scale height, but planets need to be significantly super-thermal to host classical rotationally supported CPDs. The circularization radius of inflowing gas (as a fraction of the Hill sphere radius) shows an approximately quadratic power-law scaling with the ratio of planetary mass to the thermal mass. Compared to more physically complete radiation hydrodynamic simulations, our runs almost maximize the possibility for classical CPD formation, and hence define a plausible necessary condition for CPDs. The low abundance of detected CPDs in disks where planetary companions are inferred from substructure data may be due to a combination of the large scale height of the protoplanetary disk, and a low frequency of sufficiently massive protoplanets. Unless their CPDs cool below the local protoplanetary disk temperature, most of the wide-orbit giant planet population will be embedded in quasi-spherical envelopes that are hard to detect. Disks, and satellite systems, are more likely to form around smaller orbital separation planets.
\end{abstract}


\section{Introduction} 
\label{sec:intro}
Circumplanetary disks (CPDs) are a possible byproduct of the post-runaway phase of giant planet formation \citep{Pollack1996}, when the planet contracts to a small fraction of the Hill radius while accretion from the protoplanetary disk continues. If the accreting gas has enough net angular momentum, as is very likely, and if that gas can cool sufficiently fast -- a complex and less certain proposition -- a rotationally supported disk will form. CPDs play a crucial role not only in the formation of giant planets but also in the birth of their regular satellite systems. Indeed, the planar architecture of the Galilean moons of Jupiter has long been considered as decisive evidence for a proto-Jovian CPD.

Early work on CPDs was largely motivated by the problem of satellite formation. \citet{Lunine1982} modeled the formation of the Galilean satellites in a Jovian circumplanetary disk, and understanding the environment within which satellites form has motivated subsequent one-dimensional disk models \citep{Canup2002,Mosqueira2003a,Mosqueira2003b,Martin2011,Batygin2020}. Some of these models have been adjusted to include pre-computed models for the evolution of planetary luminosity that determine the formation of icy and rocky moons around Jupiter and super-Jovian exoplanets around other stars \citep{Heller2015a,Heller2015b}. Different assumptions for feeding, angular momentum transport, and tidal truncation efficiency feed forward to the order of magnitude differences in the predicted density of these models' CPDs and uncertainty as to whether CPDs are fundamentally accretion or decretion disks.
A second motivation is exoplanetary. Circumplanetary disks can be more luminous than the planet itself, and generate distinctive kinematic signatures when observed in molecular line tracers. These properties could serve as signposts to planets that are embedded in and still accreting from the protoplanetary disk \citep{Zhu2015, Eisner2015, Perez2015, Szulagyi2019, Andrews2021}. Although there is persuasive circumstantial evidence from disk substructure observations that a large population of such planets exists \citep{Andrews2020}, most searches for CPDs have resulted in only fairly constraining upper limits \citep{Andrews2021}. A  spatially isolated CPD has been detected around the planet PDS~70~c \citep{Isella2019, Benisty2021}. Emission, consistent with a disk-embedded protoplanet at 90~AU separation from its host, has been detected for AB~Aur~b \citep{Currie2022, Zhou2022}. On large scales, the AB~Aur disk shows kinematic evidence for gravitational instability \citep{Speedie2024}. An even larger separation (200~AU) candidate has been found from molecular line imaging in the AS~209 system \citep{Bae2022}.  

Two-dimensional hydrodynamic simulations demonstrate that gas flowing toward accreting gap-opening planets has sufficient angular momentum to form CPDs \citep{Lubow1999, Dangelo2003}. In this approximation, CPDs were found to be low mass (with inferred instantaneous solid masses orders of magnitude below that of the Galilean satellites) and relatively thick structures, with ratios of vertical scale height to radius $H/R \sim$~{few}$\times 0.1$. Pioneering three-dimensional simulations of the same era, however, already revealed the necessity of the vertical dimension \citep{Dangelo2023b}. CPDs are fed with gas from out-of-plane flows, and show complex dynamics that includes meridional circulation that could not be represented in a two-dimensional approximation \citep{Tanigawa2012, Gressel2013, Szul2014, Morbidelli2014, Maeda2022}. The formation of a CPD also requires that the accreting gas be able to cool, leading to qualitatively different outcomes from isothermal versus adiabatic simulations \citep{Fung2019}. Physically, CPD formation is thus expected to depend on the opacity of the accreting material and the intrinsic luminosity of the protoplanet \citep{Ayliffe2009, Szul2016, Szul2017a, Szul2017b, Schulik2020}. Using multi-fluid radiation hydrodynamic simulations, \citet{Krapp2024} proposed that a necessary condition for CPD formation is that the cooling time be at least an order of magnitude shorter than the orbital time. Multiple other physical processes are also important. Tidal truncation efficiency varies with $H/R$ \citep{Martin2023}, magnetic fields could launch protoplanetary outflows or jets \citep{Quillen1998,Gressel2013}, while non-ideal magnetohydrodynamics may reduce the efficiency of CPD angular momentum transport \citep{Lubow2013,Fujii2017}. The transport of solid particles from the protoplanetary to the circumplanetary disk will influence the morphology of the gas flow -- primarily through dust's dominant contribution to opacity -- and will, of course, be of central importance for planetesimal and satellite formation within CPDs \citep{Maeda2024}.

The process repeats, leading to sequential satellite formation from the inside out. They found that the total mass, mass distribution, and resonant configuration of the Galilean satellites can be reproduced if the circum-Jovian disk had a dust-to-gas ratio of $\sim 0.1$ and if the migration timescale was $\sim 10^4$ years. 

The small number of detected CPDs could be reconciled with the high inferred abundance of giant planets if protoplanetary accretion, at least at large orbital radii, commonly forms quasi-spherical envelopes rather than classical CPDs. This possibility is broadly consistent with radiation hydrodynamic simluations \citep[e.g.][]{Krapp2024}, which suggest that the cooling criterion for CPD formation is hard to satisfy. Our goal in this work is to explore this hypothesis further by defining a necessary condition for CPD formation as a function of planet mass and disk aspect ratio $H/R$. We run a small grid of purely hydrodynamic simulations of planet-disk interaction, using an ``effectively isothermal" equation of state that is the most favorable for CPD formation. (As we discuss later, by not including any mass sink our runs do not strictly define a ``best case" scenario for disk formation.) We also seek to test the idea that CPD formation (with simple thermodynamics) is determined substantially by the ratio of the planetary Hill sphere to the disk scale height (i.e. whether the planet is sub- or super-thermal).

Our paper is structured as follows: \S\ref{sec:methods} presents the relevant equations and simulation setup; \S\ref{sec:results} reports our findings; \S\ref{sec:discussion} analyzes these results; and \S\ref{sec:conclusions} offers concluding remarks.

\section{Methods}
\label{sec:methods}

We perform three-dimensional hydrodynamic simulations of circumplanetary disks using the grid-based code \texttt{Athena++} \citep{Stone2020}. The simulations are carried out in spherical polar coordinates $(r, \theta, \phi)$, where $r$ is the radial distance, $\theta$ is the polar angle, and $\phi$ is the azimuthal angle.

\subsection{Disk model}
The simulations solve the compressible Euler equations:
\begin{equation}
\frac{\partial \rho}{\partial t} + \nabla \cdot (\rho \mathbf{v}) = 0,
\end{equation}
\begin{equation}
\label{eq:momentum}
\frac{\partial (\rho \mathbf{v})}{\partial t} + \nabla \cdot (\rho \mathbf{v} \otimes \mathbf{v} + P\mathbf{I}) = -\rho \nabla \Phi, 
\end{equation}
\begin{equation}
\label{eq:energy}
\frac{\partial E}{\partial t} + \nabla \cdot [(E + P)\mathbf{v}] = -\rho \mathbf{v} \cdot \nabla \Phi, 
\end{equation}
where $\rho$ is the density, $P$ the gas pressure, $\mathbf{v}$ the velocity, and $\Phi$ the gravitational potential of the central star and the planet, and $E = u+\lvert \mathbf{v} \rvert^2 / 2$ is the specific total (internal + kinetic) energy.

The simulations are performed in a frame centered on the star, with the planet in a fixed circular orbit in the disk midplane. We set $GM_\star = 1$ and the radial location of the planet, $R_p = 1$, so that the Keplerian velocity $v_k = \sqrt{GM/r}$ and frequency $\Omega_k = \sqrt{GM/r^3}$ both equal 1 at the planet's orbit, therefore we define the orbital frequency at planet's location as $\Omega_p = 1$.
\begin{equation}\label{eq:potential}
\begin{aligned}
     \Phi = & -\frac{G(M_*+M_p)}{1+q} \bigg[\frac{1}{r} \\
     & + \frac{q}{\sqrt{r^2+R_p^2-2rR_p\cos{\phi'}+\epsilon^2}} - \frac{qR\cos{\phi'}}{R_p^2}\bigg] \\
\end{aligned}
\end{equation}
In Equation \ref{eq:potential}, $q = M_p/M_*$ is the mass ratio of a protoplanet to a star, $\epsilon$ is the smoothing length of the planet's potential which is equal to $0.012 R_p$, and $\phi'= \phi-\phi_p$ denotes the azimuthal separation from the planet.

\begin{figure}[hbt!]
% \vspace{-2cm}
    \includegraphics[width=0.46\textwidth]{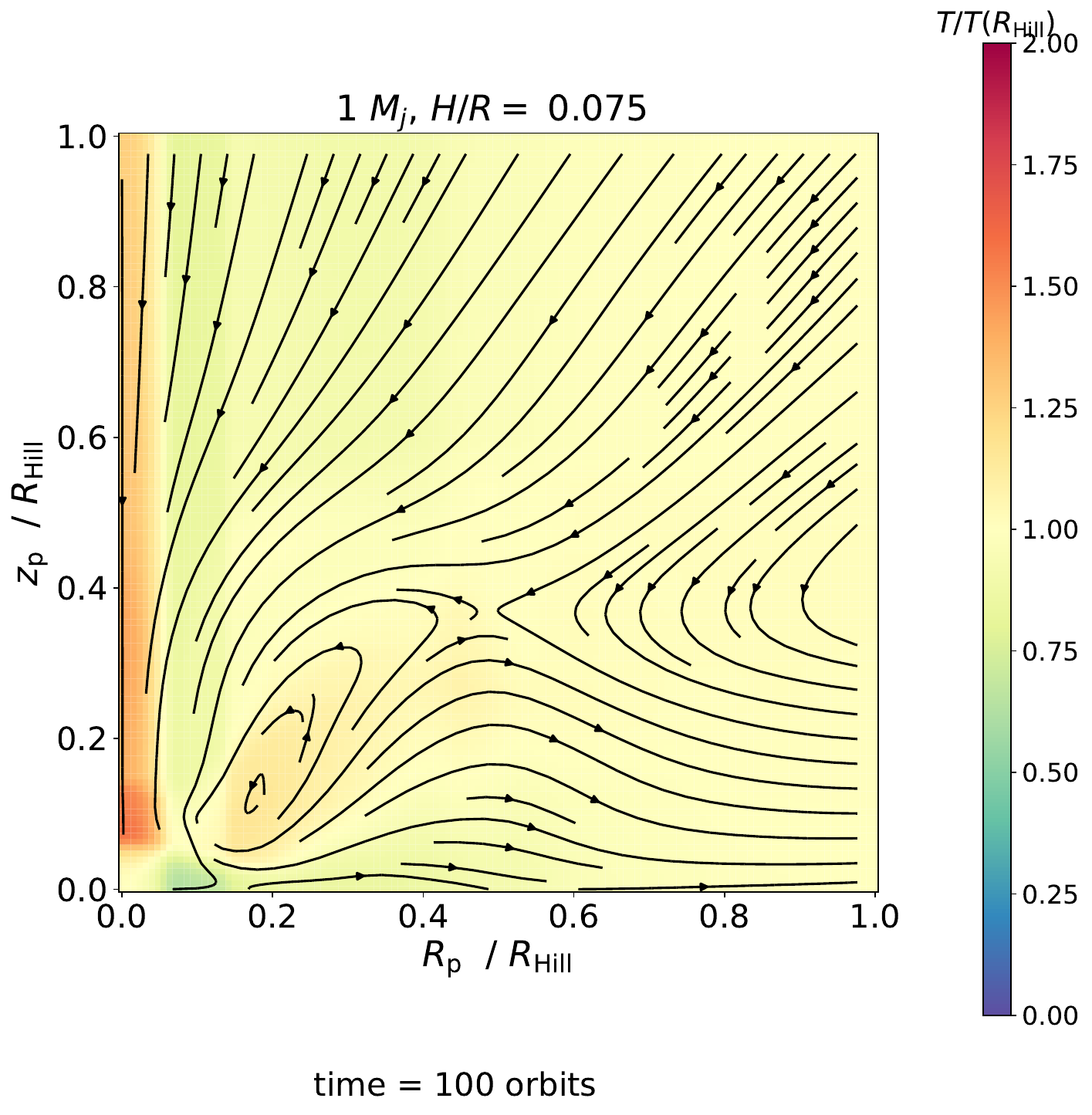}
    % \vspace{-3cm}
        \caption{Vertical slice through a CPD for 1 Jupiter-mass and disk aspect ratio of 0.075 after 100 orbital periods. The figure shows the temperature ratio at a certain radius to the temperature at a 1 Hill radius. The x-axis represents the radial distance from the planet normalized by the Hill radius, while the y-axis shows the vertical distance.}
        \label{fig:temperatures}
\end{figure}

A key parameter in our simulations is the Hill radius (of a Hill sphere), which represents the approximate spherical region around a planet within which the planet's gravitational influence dominates over that of its parent star. It is defined as:
\begin{equation}\label{eq:Hill}
    R_{\rm Hill}=R_{\rm p}\left(\frac{q}{3}\right)^{\frac{1}{3}}
\end{equation}
%
% We use three different planetary masses: $q=0.9\times 10^{-3}$ (1 Jupiter mass), $q=0.27\times 10^{-3}$ (0.3 Jupiter masses), and $q=2.7\times 10^{-3}$ (3 Jupiter masses). 
We employ an equation of state given as:
\begin{equation}\label{eq:pressure}
    p = \frac{c_s^2\rho_0}{\gamma}\left(\frac{\rho}{\rho_0}\right)^\gamma,
\end{equation}
where 
\begin{equation}\label{eq:pressure-1}
    p = u\rho(\gamma -1),
\end{equation}
but we set $\gamma = 1.01$. This choice of $\gamma$ is motivated by our desire to model a disk that is very close to isothermal conditions while avoiding the numerical challenges associated with a \textit{strictly} isothermal equation of state ($\gamma = 1$). Our initial attempts to simulate the system with a pure isothermal setup resulted in numerical instabilities and convergence issues, particularly in regions of strong density gradients. Notably, isothermal CPDs maintain a disk-like structure, whereas adiabatic CPDs tend to have a more spherical shape \citep{Fung2019, Miranda2020}. Figure \ref{fig:temperatures} illustrates the ratio of temperatures at various radii to the temperature at one Hill radius. We see minimal temperature variations across the disk, though there are larger temperature perturbations in a narrow circumpolar region above and below the planet.

In our disk model, we use the relation

\begin{equation}\label{eq:sound-speed}
    c_s = \frac{H}{R}v_k
\end{equation}
to set the initial condition for the sound speed, which allows us to express the sound speed in terms of the disk's \textit{initial} aspect ratio ($H/R$). Equation \ref{eq:sound-speed} follows from vertical hydrostatic equilibrium. The ratio of the vertical height $H$ to the radial distance $R$ is taken as a fixed input parameter. During the simulation, the sound speed evolves according to our hydrodynamical equations. We refer to the initially specified $H/R$ as a label for the different models.

Another important parameter in our simulations is the thermal mass ratio, often denoted as $q_{\rm thermal}$, which represents the mass at which a planet's Hill radius becomes comparable to the local scale height of the protoplanetary disk. More specifically:
\begin{equation}\label{eq:qthermal}
    q_{\rm thermal} = (H/R)^3 M_\star
\end{equation}
Here, we mostly utilize the ratio $q/q_{\rm thermal}$, which represents the ratio between the planetary physical mass and its thermal mass, and the star's physical mass is set to $M_\star=1$. 

We also define the angular momentum flux $\mathcal{\dot{L}}$ as the rate at which angular momentum is transferred through a surface (per unit area) by the flowing gas. Using this angular momentum flux and the mass flux $\mathcal{\dot{M}}$, we can define the centrifugal or \textit{circularization} radius:
\begin{equation}\label{eq:rcirc}
    R_{\rm circ} = \frac{\mathcal{\dot{L}}^2}{\mathcal{\dot{M}}^2 GM_p},
\end{equation}
where $\mathcal{\dot{L}}$ denotes the angular momentum flux, $\mathcal{\dot{M}}$ is the mass flux, $G$ is the gravitational constant, and $M_p$ is the mass of the protoplanet. In our calculations, we consider only the inflowing material when determining $R_{\rm circ}$. The circularization radius marks the transition point where the rotational support of the infalling material becomes significant compared to its radial infall velocity. Material with sufficient angular momentum to reach this radius is likely to form a disk, while material with less angular momentum will fall closer to the protoplanet before being incorporated into the disk. 

Our simulations span 100 planetary orbits. We observe that quasi-steady states are generally achieved before reaching 100 orbits, and therefore, the key characteristics of our results are already well-established at 100 orbits. 
All reported results represent averages taken over the final orbital period of each simulation. This approach allows us to present data that best represents the typical state of the system, minimizing the impact of short-term oscillations.

While our simulations are conducted in a star-centered spherical grid, for analytical purposes, we also employ a planet-centered cylindrical coordinate system. In the star-centered system, we use $\{R, \theta, \phi\}$ to represent the radial, polar, and azimuthal coordinates, respectively. We utilize $\{r, \varphi, z\}$ for the planet-centered system to represent the radial, azimuthal, and vertical coordinates, respectively. This transformation allows for a more intuitive interpretation of the results.

For our parameter study, we use three different planetary masses: $q=0.9\times 10^{-3}$ (1 Jupiter mass), $q=0.27\times 10^{-3}$ (0.3 Jupiter masses), and $q=2.7\times 10^{-3}$ (3 Jupiter masses). We also vary $c_s$ to be either $0.05R_p\Omega_p$, $0.075R_p\Omega_p$, and $0.1R_p\Omega_p$, which respectively correspond to disk aspect ratios $H/R$ of 0.05, 0.075, and 0.1 at the planet's location (see Table \ref{tab:models}).

\subsection{Initial and Boundary conditions}
% In our initial simulations, we observed that even after 100 orbital periods, the disk failed to develop a gap around the planet's orbit naturally. This is likely due to the short timescales we can simulate given computational constraints. Gap opening typically occurs on longer timescales, especially for lower-mass planets.
Planets open gaps in both inviscid and viscous disks, with a time scale that depends on planet mass and disk viscosity. In preliminary runs, we found that, starting from an unperturbed disk profile, 100 orbital periods was not sufficient to reach an approximate steady-state for the gap profile. To address this, we implement an artificial gap profile in our initial conditions to address this limitation and study the dynamics of CPDs in a more realistic context. While this approach doesn't capture the full gap-opening process, it reasonably approximates the expected disk structure for a system that has evolved for many more orbits than we can feasibly simulate. This method is similar to approaches used in other studies \citep[e.g.,][]{Kley1999, Tanigawa2012} to isolate and study specific aspects of planet-disk interactions.
The disks are initialized in hydrostatic equilibrium with a density profile:

\begin{equation}\label{eq:density}
\begin{aligned}
     \rho = & \rho_0 \bigg[\left(\frac{R}{R_p}\right)^{(-\beta+\frac{3}{2})\frac{2(\gamma-1)}{\gamma+1}} \\ 
    & - \frac{GM(\gamma-1)}{c_0^2}\left(\frac{1}{R}-\frac{1}{r}\right)\bigg]^{\frac{1}{\gamma-1}} \\
    & \times \left(1 - \exp{-\frac{(R-R_p)^2}{(R_p\Delta r)^2}}\right)
\end{aligned}
\end{equation}
$\beta=\frac{7}{3}$ defines the surface density profile $\Sigma\propto R^{-\beta}$. Here, $\Delta r = 0.1$ is the fraction that multiplied to $R_p$ gives $R_p\Delta r$, the depth of the gap.

Sometimes, we utilize the surface density $\Sigma = \sqrt{2\pi H}\rho$. The density profile in our model results in a surface density profile of  $\Sigma\propto R^{-7/3}$, which is steeper than the commonly used minimum mass solar nebula model ($\Sigma\propto R^{-3/2}$). This steep profile is a consequence of our chosen initial conditions for hydrostatic equilibrium, specifically our global isothermal equation of state with $\gamma = 1.01$. While steeper than some standard models, similar profiles have been suggested in certain contexts of planet formation and disk evolution (e.g., \cite{Chambers2009, Raymond2014}).

This profile represents a compact, rapidly decreasing density distribution around the planet, which may be appropriate for young, still-forming CPDs. Such a steep profile could result from the initial stages of disk formation, where material is concentrated close to the planet. 

The radial pressure gradient modifies the orbital frequency of the disk in hydrostatic equilibrium:
\begin{equation}
\Omega = \sqrt{\Omega_k^2 + \frac{1}{r\rho}\frac{\partial p}{\partial r}},
\end{equation}
while the initial radial and polar velocities are zero.

In the ghost zones of the inner boundary, density, and velocity are determined by the steady-state solution used for initial conditions. At the outer radial boundary, we apply outflow boundary conditions to fluid variables.

\subsection{Grid setup}
We use the spherical-polar grid with a nested static mesh refinement (SMR) around the planet's location. The intervals that span in radial, polar, and azimuthal directions are $[0.4, 2.5]$, $[9\pi/20, 11\pi/20]$, and $[0, 2\pi]$, respectively. Aiming for $\sim 50$ cells around the Hill sphere, we come up with the effective resolution at the \emph{root} level: $168$ cells across the $r$-direction, $576$ cells across the $\phi$-direction, and $72$ cells across the $\theta$-direction for $H/R = 0.05$. For higher scale heights, we increase the $\theta$ domain to accommodate the increased vertical extent of the disk. Consequently, for $H/R = 0.075$ and $H/R = 0.1$, the number of cells across the $\theta$-direction is increased to 120 and 168, respectively (see Table \ref{tab:models}).

The nested SMR helps us lower the resolution at the root level, keeping a high enough resolution around the Hill 
sphere to resolve the CPD properly. The detailed descriptions of all the refined regions are shown in Table \ref{tab:smr}.

\begin{table*}[]
\centering
\caption{Refinement Domain}
\begin{tabular}{lcccc}
\hline
Refinement Level & Domain Boundaries & \multicolumn{3}{c}{Number of Cells ($r \times \phi \times \theta$)} \\
 & & H/R = 0.05 & H/R = 0.075 & H/R = 0.1 \\
\hline
Root Level & Full domain & $168 \times 576 \times 72$ & $168 \times 576 \times 120$ & $168 \times 576 \times 168$ \\
\hline
Refinement Level 1 & $9\pi/20 \rightarrow 11\pi/20 \; \theta$ & & & \\
 & $0.8 \rightarrow 1.2 \; r$ & $64 \times 73 \times 144$ & $64 \times 73 \times 240$ & $64 \times 73 \times 336$ \\
 & $\pm 0.2 \; \phi$ & & & \\
\hline
Refinement Level 2 & $19\pi/40 \rightarrow 21\pi/40 \; \theta$ & & & \\
 & $0.9 \rightarrow 1.1 \; r$ & $64 \times 73 \times 144$ & $64 \times 73 \times 240$ & $64 \times 73 \times 336$ \\
 & $\pm 0.1 \; \phi$ & & & \\
\hline
Refinement Level 3 & $39\pi/80 \rightarrow 41\pi/80 \; \theta$ & & & \\
 & $0.95 \rightarrow 1.05 \; r$ & $64 \times 73 \times 144$ & $64 \times 73 \times 240$ & $64 \times 73 \times 336$ \\
 & $\pm 0.05 \; \phi$ & & & \\
\hline
\end{tabular}
\label{tab:smr}
\end{table*}

\begin{table}
\centering
\caption{Model Parameters}
\label{tab:models}
\begin{tabular}{ccccccc}
\hline \hline
Model ID & $H/R$ & Resolution & $M_p$ & $q_\mathrm{thermal}$ & $\theta$-domain \\
 & & (cells/$r_H$) & $(M_J)$ & & (\# of cells)  \\
\hline
1m0.05h & 0.05 & 42 & 1 & 7.2 & 72 \\
0.3m0.05h & 0.05 & 28 & 0.3 & 2.2 & 72 \\
3m0.05h & 0.05 & 61 & 3 & 21.6 & 72 \\
1m0.075h & 0.075 & 42 & 1 & 2.1 & 120 \\
03m0.075h & 0.075 & 28 & 0.3 & 0.6 & 120 \\
3m0.075h & 0.075 & 61 & 3 & 6.4 & 120 \\
1m0.1h & 0.1 & 42 & 1 & 0.9 & 168 \\
03m0.1h & 0.1 & 28 & 0.3 & 0.3 & 168 \\
3m0.1h & 0.1 & 61 & 3 & 2.7 & 168 \\
1m0.05h-HR & 0.05 & 67 & 1 & 7.2 & 120 \\
\hline

\end{tabular}
\end{table}

\section{Results}
\label{sec:results}
\begin{figure*}[hbt!]
    \center
    \includegraphics[width=0.9\textwidth]{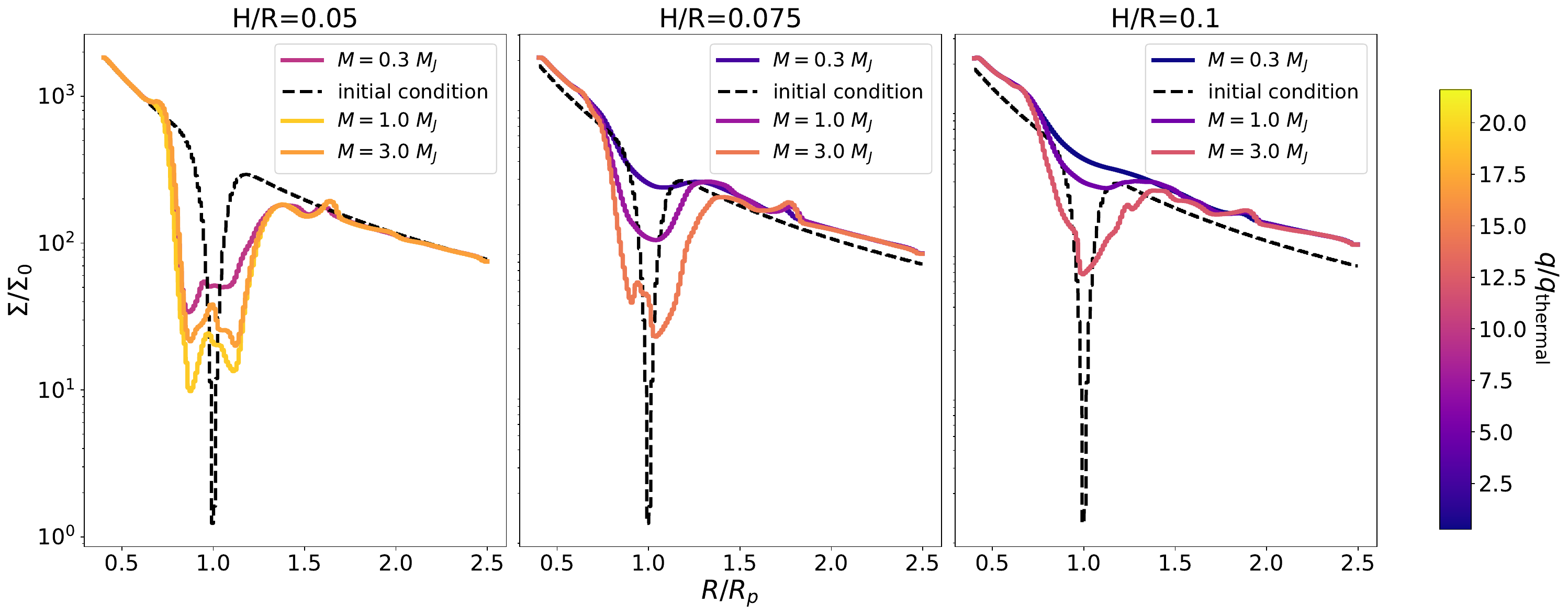}
        \caption{Gap profiles for various disk and planet configurations. Each line shows the azimuthally averaged surface density $\Sigma$ normalized to the initial surface density at the planet's orbital radius $\Sigma_0$, plotted against radial distance from the star normalized to the planet's orbital radius $R/R_p$. Different colors represent different models. Note the deeper gaps for lower $H/R$ values and higher planet masses, and the shallower, less distinct gaps for higher $H/R$ values and lower planet masses. All profiles are time-averaged over the final orbital period of the simulation.}
        \label{fig:gap-profiles}
\end{figure*}

In this section, we present the key findings from our three-dimensional hydrodynamic simulations of planet-disk interactions, focusing on the formation and structure of CPDs and envelopes. Our simulations explore a range of planet masses (0.3, 1, and 3 Jupiter masses) and disk aspect ratios ($H/R$ = 0.05, 0.075, and 0.1), allowing us to investigate how these parameters influence the dynamics and morphology of the circumplanetary environment. We begin by examining the gap profiles and midplane density distributions to establish the overall impact of the planet on the protoplanetary disk.
Figures \ref{fig:gap-profiles} and \ref{fig:midplane-cuts} illustrate the trends in disk structures and flow patterns across different planetary masses and disk aspect ratios. These trends align well with both theoretical expectations (e.g., \citealt{Crida2006}) and previous numerical studies (e.g., \citealt{Fung2014}), demonstrating the consistency and accuracy of our simulations.

\subsection{Global gas profiles}
\begin{figure*}[hbt!]
% \vspace{-2cm}
    \includegraphics[width=\textwidth]{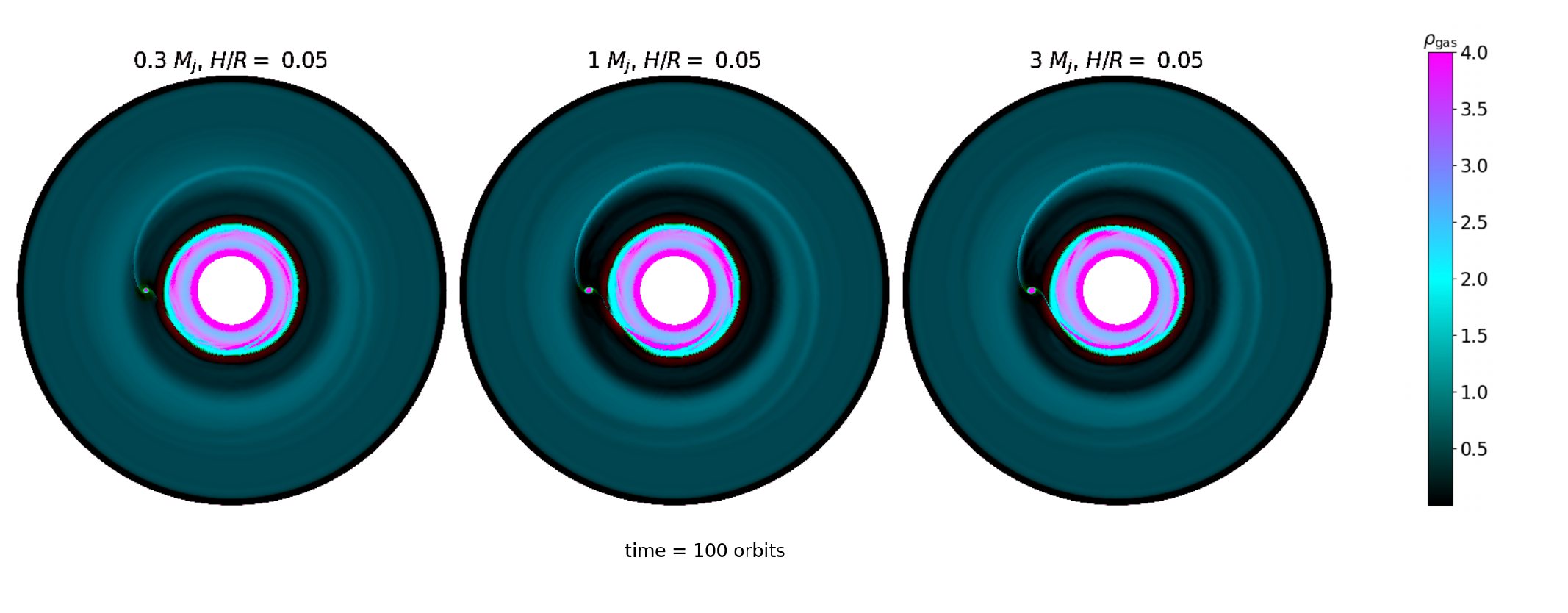}
    % \vspace{-3cm}
        \caption{Mid-plane gas density distributions for different planet masses at 0.075 aspect ratio after 100 orbital periods. Each panel shows a face-on view of the disk, with the planet located at the left part of the disk. Columns show different planet masses: 0.3 $M_J$ (left), 1 $M_J$ (center), and 3 $M_J$ (right). The color scale indicates gas density, ranging from low (dark) to high (bright) on a logarithmic scale. Note the variations in gap depth and spiral wave patterns across different models.}
        \label{fig:midplane-cuts}
\end{figure*}

In disks with higher scale heights, we observe shallower and less pronounced gaps. This is particularly evident when the disk's aspect ratio exceeds the planet's Hill radius. In these cases, the disk's thermal pressure effectively resists the planet's gravitational influence, preventing the formation of a deep gap. Conversely, simulations with smaller scale heights ($H/R < R_{\rm Hill}$) exhibit deeper and more defined gaps. Here, the planet's gravitational perturbations dominate over the disk's pressure support, efficiently clearing material from its orbital vicinity. We can also observe these effects in the midplane density distributions (Figure \ref{fig:midplane-cuts}). Figure \ref{fig:midplane-cuts} presents mid-plane density distributions for our suite of simulations after 100 orbital periods. The columns represent different planet masses: 0.3 Jupiter masses (left), 1 Jupiter mass (center), and 3 Jupiter masses (right). There's a progression from lower-mass planets to higher-mass planets that shows increasingly pronounced gaps. This progression highlights the balance between planetary gravity and disk pressure in shaping the surrounding environment.
\subsection{Circumplanetary gas profiles}
\subsubsection{Vertical structure}
\begin{figure*}[hbt!]
% \vspace{-2cm}
    \includegraphics[width=\textwidth]{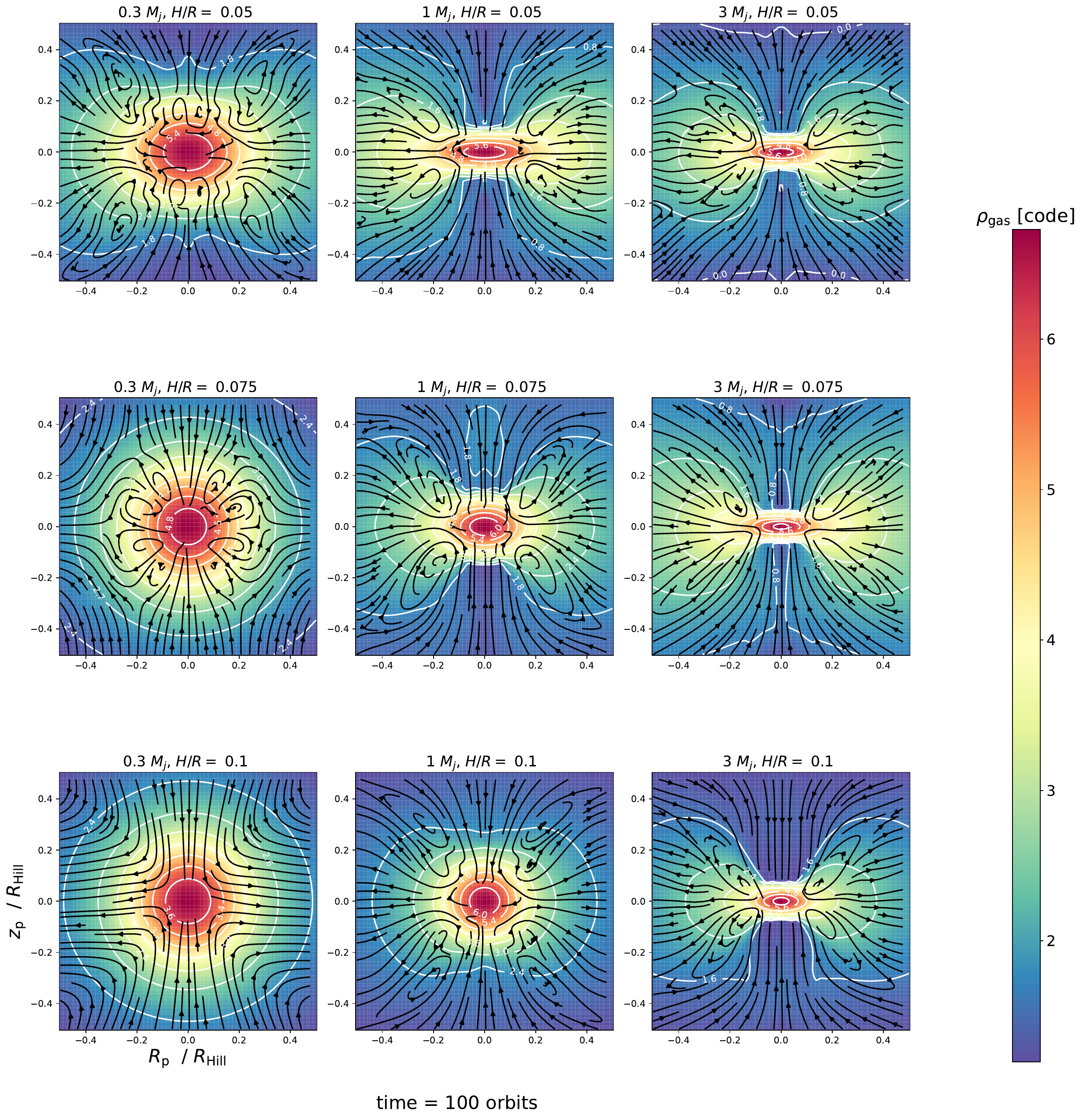}
    % \vspace{-3cm}
        \caption{Vertical slices through CPDs for various planet masses and disk aspect ratios after 100 orbital periods. Each panel shows the gas density (color scale, in code units) and velocity streamlines in a region centered on the planet. The x-axis represents the radial distance from the planet normalized by the Hill radius, while the y-axis shows the vertical distance. Rows correspond to different disk aspect ratios: $H/R = 0.05$ (top), 0.075 (middle), and 0.1 (bottom). Columns represent different planet masses: 0.3 $M_J$ (left), 1 $M_J$ (center), and 3 $M_J$ (right), where $M_J$ is Jupiter's mass. The streamlines illustrate the complex flow patterns around the planet, including accretion, circulation within the CPD, and outflows. Note the increasing prominence and vertical extent of the CPDs with increasing planet mass and disk aspect ratio.}
        \label{fig:vertical-cuts}
\end{figure*}

We also examined the vertical structure of the circumplanetary environment. Figure \ref{fig:vertical-cuts} illustrates the complex gas dynamics around planets of varying masses embedded in disks with different aspect ratios. This figure showcases vertical slices through the disk, centered on the planet. The radial and vertical distances are normalized by the Hill radius, with gas density represented by the color scale and velocity fields depicted through streamlines.

As we examine the panels from left to right, increasing in planetary mass, we observe a clear progression in the prominence and extent of the circumplanetary disks. The most massive planets, represented in the rightmost column, exhibit well-defined disk structures that extend further both radially and vertically compared to their lower-mass counterparts. 
There is also a variation in disk structure from top to bottom, corresponding to increasing disk aspect ratios. Higher $H/R$ values result in thicker, more vertically extended CPDs, reflecting the influence of the background disk's scale height on the gas dynamics around the planet. For the lower-mass planets, particularly the 0.3 Jupiter mass cases, we observe that a distinct CPD does not form. Instead, these planets are surrounded by a more spherical envelope of gas. This envelope lacks the clear rotational structure seen around more massive planets and appears to be more directly influenced by the background disk's properties. These trends are consistent with an extrapolation of the analytic expectation for the scale height $h/r$ of a thin CPD, which at a fixed fraction of the Hill sphere radius is given by,
\begin{equation}
   \left( \frac{h}{r} \right) \propto \left( \frac{H}{R} \right) q^{-1/3},
\end{equation}
where $H/R$ is the protoplanetary disk scale height.

We also observe material accreting onto the disk from higher altitudes, followed by complex circulation within the disk itself. Notably, many cases, particularly those involving higher-mass planets, show evidence of outflow in the midplane regions. This suggests a dynamic exchange of material between the CPD and the surrounding environment -- i.e.  decretion. 

\begin{figure*}[hbt!]
% \vspace{-2cm}
    \includegraphics[width=\textwidth]{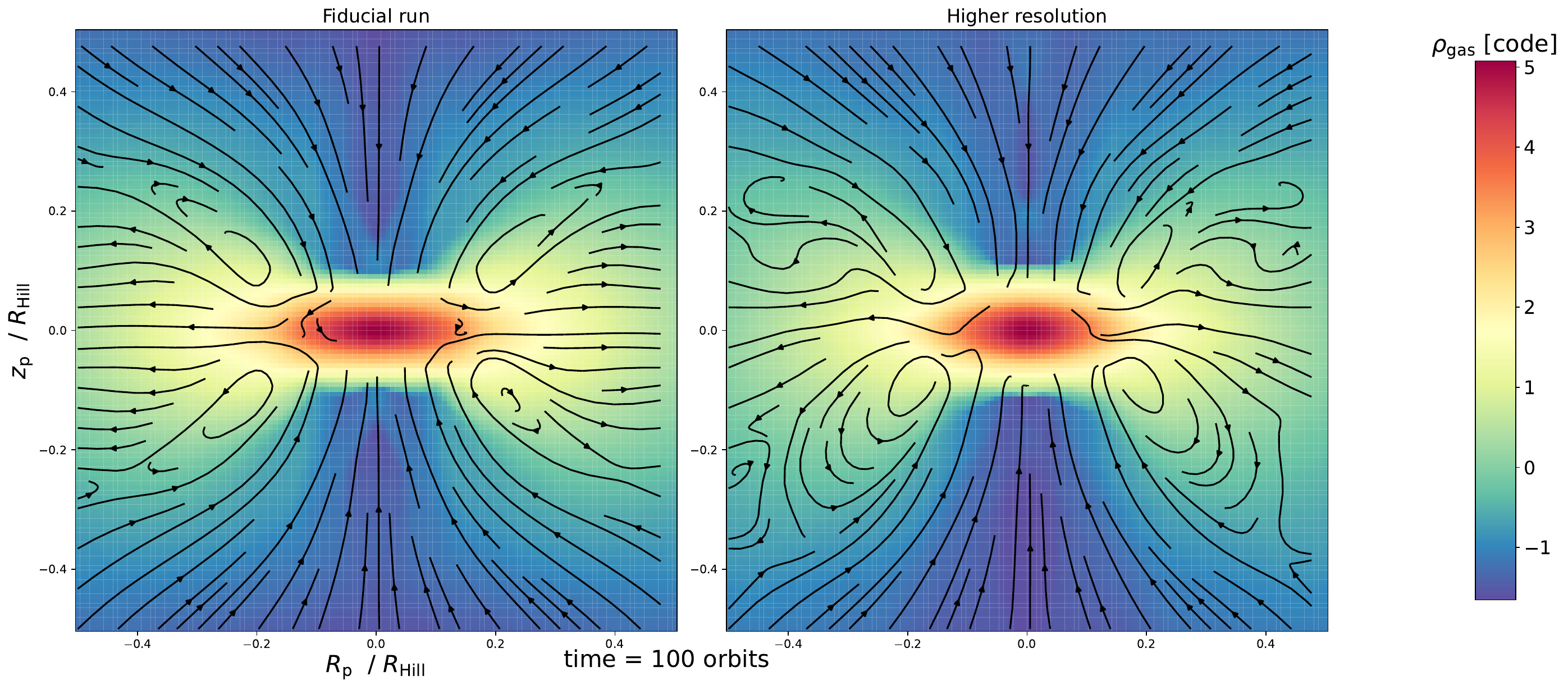}
    % \vspace{-3cm}
        \caption{Comparison of gas density distributions in CPDs between the fiducial run (left panel) and a higher resolution simulation (right panel) at 100 orbital periods. The color scale represents gas density in code units. Both axes are normalized by the Hill radius ($R_{\rm Hill}$). The x-axis shows the radial distance from the planet ($R_p$), while the y-axis represents the vertical distance ($z_p$). Both simulations demonstrate similar disk structures extending to approximately 0.4 $R_{\rm Hill}$ radially and 0.2 $R_{\rm Hill}$ vertically, with the fiducial run exhibiting a slightly more pronounced disk-like morphology.}
        \label{fig:high-res}
\end{figure*}

To see if any of our results are influenced by insufficient resolution, we did a run with a higher resolution, \texttt{1m0.05h-HR} in Table \ref{tab:models}. Figure \ref{fig:high-res} presents a comparison between our fiducial run (1 Jupiter mass and $H/R=0.05$) and a higher resolution simulation (denoted as \texttt{1m0.05h-HR} in Table \ref{tab:models}) of the CPD at 100 orbital periods. The figure displays the gas density distribution in the $R-z$ plane, where R and z are normalized by the Hill radius ($R_{\rm Hill}$). 

We observe striking similarities between the two simulations, with only minor differences in the fine structure of the disk. Both runs exhibit a flattened, disk-like structure around the central planet, with the highest densities concentrated near the midplane and decreasing radially outward, though the fiducial run appears to display a slightly more pronounced disk-like morphology compared to the higher resolution simulation. This is evidenced by a more distinct flattening of the high-density region (red and yellow areas) along the midplane in the fiducial run. The higher resolution run, while broadly similar, shows a marginally more vertically extended structure.
This result strongly suggests that our fiducial resolution is sufficient to capture the essential dynamics and structure of the CPD. The minor differences observed do not significantly alter the overall conclusions drawn from our simulations. 

\subsubsection{Midplane flows}
\begin{figure*}[hbt!]
% \vspace{-2cm}
    \includegraphics[width=\textwidth]{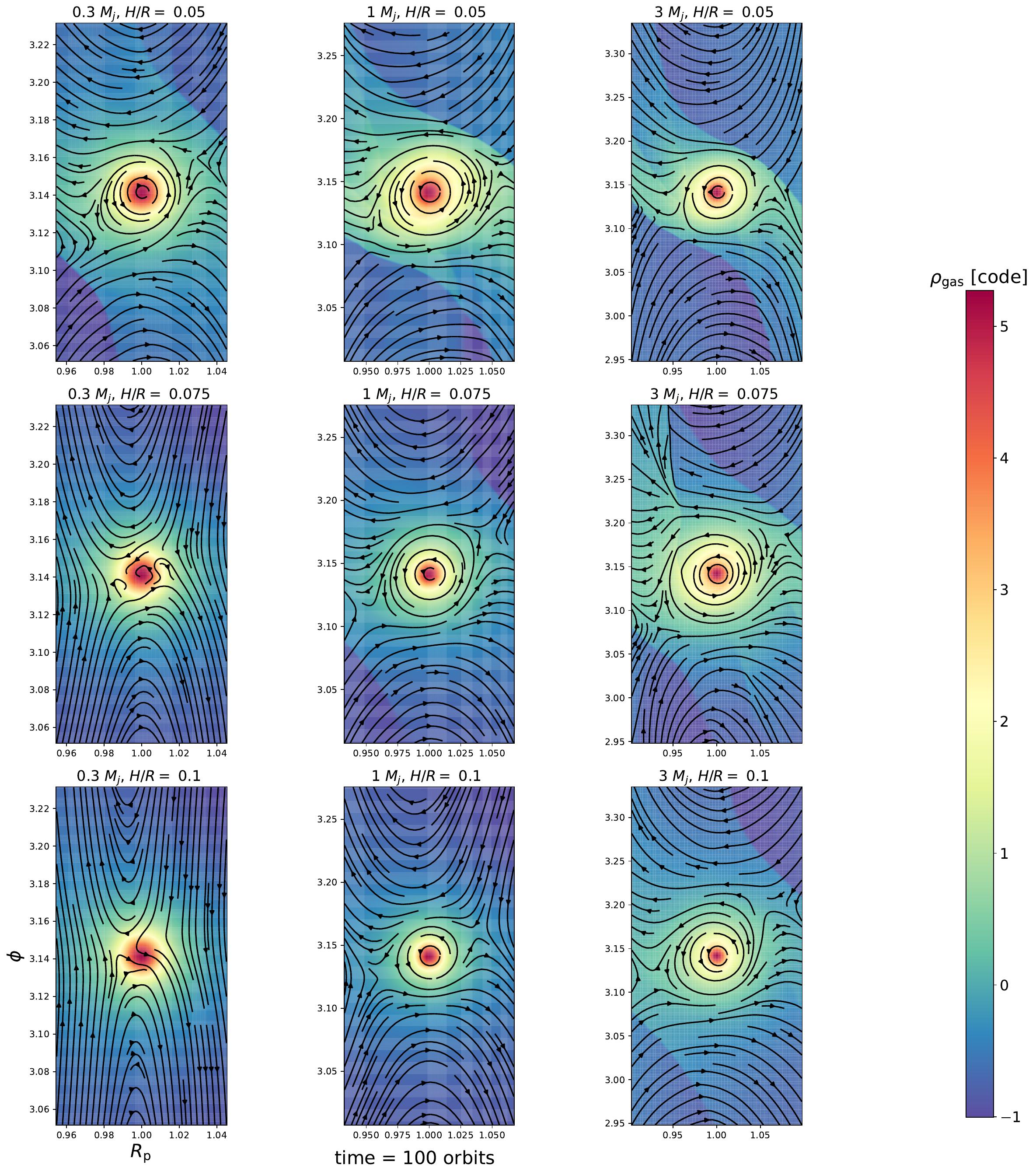}
    % \vspace{-3cm}
        \caption{Midplane slices through CPDs for various planet masses and disk aspect ratios after 100 orbital periods. Each panel shows the gas density (color scale, in code units) and velocity streamlines in a region centered on the planet. The x-axis represents the radial distance from the planet spanning 1 Hill radius, while the y-axis shows the azimuthal direction. Rows correspond to different disk aspect ratios: $H/R = 0.05$ (top), 0.075 (middle), and 0.1 (bottom). Columns represent different planet masses: 0.3 $M_J$ (left), 1 $M_J$ (center), and 3 $M_J$ (right), where $M_J$ is Jupiter's mass. The streamlines illustrate the complex flow patterns around the planet, including accretion, circulation within the CPD, and horseshoe region.}
        \label{fig:spirals}
\end{figure*}

Having confirmed the numerical consistency of our simulations through resolution testing, we also examine the structural characteristics of CPDs on the mid-plane across the full range of planetary masses and disk aspect ratios investigated in this study. Figure \ref{fig:spirals} provides a comprehensive view of CPD structures across various planetary masses and disk aspect ratios. The rows represent different disk aspect ratios ($H/R =$ 0.05, 0.075, and 0.1), and columns depict planetary masses of 0.3, 1, and 3 Jupiter masses ($M_J$). This layout facilitates a direct comparison of CPD structures across our parameter space. Each panel displays the gas density distribution in the disk midplane, with the color scale representing density in code units.

As we examine the panels, just like in Figure \ref{fig:vertical-cuts}, we observe that planetary mass plays a crucial role in determining the CPD's size and structure. More massive planets, with their stronger gravitational fields, tend to accumulate larger, more extensive disks. This is particularly evident in the rightmost column (3 $M_J$), where the CPDs exhibit a broader radial extent compared to their lower-mass counterparts.

The disk aspect ratio also significantly influences CPD morphology. Higher aspect ratios, indicative of hotter or more vertically extended disks, appear to result in less sharply defined structures. This is also seen in Figure \ref{fig:vertical-cuts}.

The spiral structures are visible in many of the panels, particularly for the higher-mass planets. These spirals are a manifestation of density waves, excited by the gravitational interaction between the planet and its disk, which we also observe in Figure \ref{fig:midplane-cuts}. The clarity and strength of these spirals vary with both planetary mass and disk aspect ratio, reflecting the complex balance between gravitational, pressure, and rotational forces within the CPD. Figure \ref{fig:spirals} showcases the diversity of CPD structures that may exist in nature and highlights the importance of considering both planetary and disk properties in understanding CPD formation and evolution. 

\begin{figure*}[hbt!]
% \vspace{-2cm}
    \center
    \includegraphics[width=0.9\textwidth]{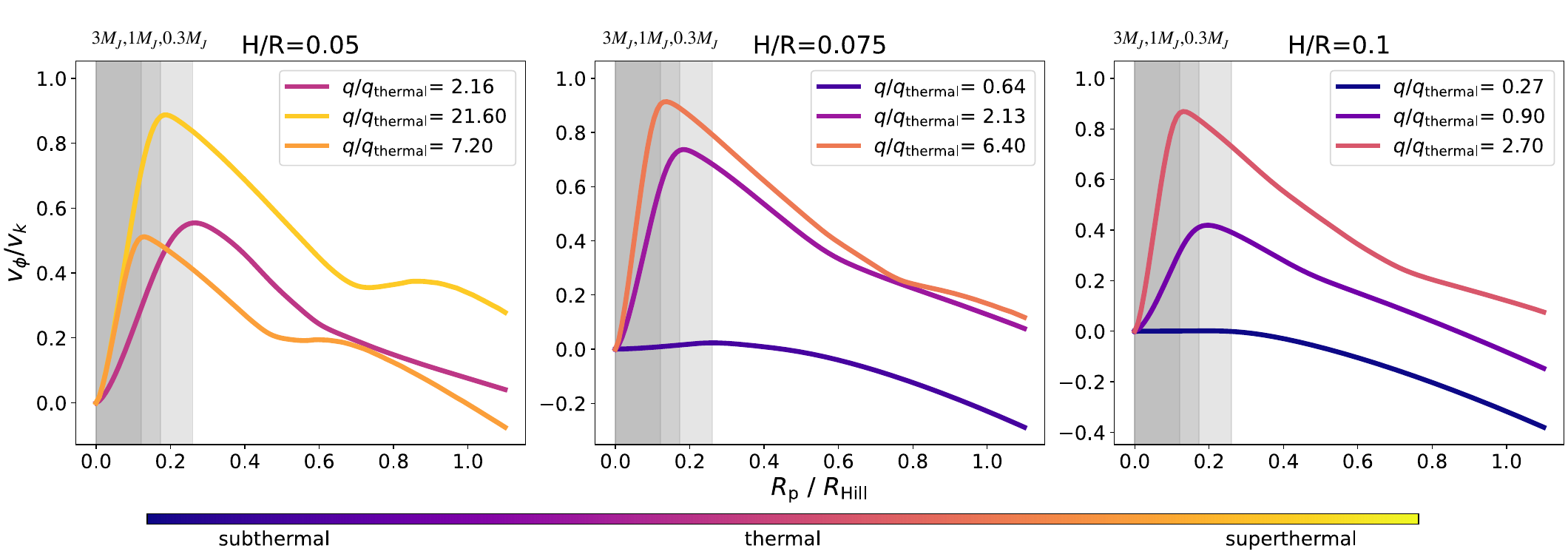}
    % \vspace{-3cm}
        \caption{Azimuthal velocity profiles around planets of different masses in disks with varying aspect ratios. The x-axis shows the radial distance from the planet normalized by the Hill radius, while the y-axis represents the ratio of local azimuthal velocity to Keplerian velocity $\sqrt{GM_p/r}$. Different colors indicate various combinations of planet mass (0.3, 1, and 3 Jupiter masses) and disk aspect ratios ($H/R$ = 0.05, 0.075, and 0.1). The sub-Keplerian nature of the flow is evident across all cases. The grey region is the region below the smoothing length, which changes for a different planet mass. We show which planet mass the region corresponds to above the panels.}
        \label{fig:keplerian}
\end{figure*}

To better understand the rotational behavior of these circumplanetary structures, we analyzed the azimuthal velocity profiles presented in Figure \ref{fig:keplerian}. Figure \ref{fig:keplerian} presents a visualization of the azimuthal velocity profiles in the vicinity of planets with varying masses embedded in disks of different aspect ratios. The x-axis represents the radial distance from the planet normalized by the Hill radius, while the y-axis shows the ratio of the local azimuthal velocity to the Keplerian velocity at that radius. One of the most striking features of this plot is the consistent sub-Keplerian rotation observed in all cases. This sub-Keplerian rotation is characteristic of pressure-supported disks and envelopes, where the radial pressure gradient provides partial support against the planet's gravity, reducing the need for centrifugal balance. We observe a clear mass dependence in the velocity profiles. The higher-mass planets (3 Jupiter masses) exhibit more superthermal regimes and velocity profiles that approach Keplerian rotation more closely, especially at larger radii. This suggests that the gravitational influence dominates over pressure effects for more massive planets, leading to a more disk-like rotational structure. In contrast, the lower mass planets (0.3 Jupiter masses) show significantly sub-Keplerian rotation, indicative of a more pressure-supported, envelope-like structure, as seen in Figure \ref{fig:vertical-cuts}. The disk aspect ratio also plays a role in shaping these velocity profiles. For a given planetary mass, higher aspect ratios ($H/R$ = 0.1) generally result in more sub-Keplerian rotation compared to lower aspect ratios ($H/R$ = 0.05). This is consistent with the idea that thicker disks have stronger pressure support, leading to greater deviations from Keplerian rotation. The convergence of the velocity profiles at larger radii, particularly for the higher mass planets, suggests that the rotational dynamics beyond about $0.8-1.0$ $R_{\rm Hill}$ are increasingly dominated by the background disk rather than the planet's local influence. The long-term evolution of the planet-disk system suggests that there is ongoing angular momentum exchange between the circumplanetary material and the larger protoplanetary disk. This smoothing length appears as different physical scales when normalized by the Hill radius for different planet masses, which explains the consistent appearance of the grey regions across all panels. In our simulations, the smoothing length ($\epsilon$ in Equation \ref{eq:potential}) ranges from 0.12 to 0.27 Hill radii. This choice of smoothing length has important implications for the flow structure near the planet, potentially affecting the region up to several times $\epsilon$, which constitutes a significant fraction of the Hill sphere. The relatively large smoothing length may enhance pressure support close to the planet, potentially favoring the formation of an envelope-like structure rather than a distinct disk. This demonstrates the delicate balance between computational feasibility and physical accuracy in such simulations, as reducing the smoothing length would allow for more detailed resolution of the inner regions but would significantly increase computational costs. However, we have confirmed that even in cases where the smoothing length is the same across different planetary masses, the fundamental transition from envelope-dominated to disk-dominated structures remains primarily governed by physical parameters (planet mass and disk aspect ratio) rather than numerical artifacts.

When considering the gravitational potential around planets in numerical simulations, the choice of smoothing length plays a crucial role in accurately modeling circumplanetary structures. Recent work by \cite{Schulik2020} demonstrates that the smoothing length must be resolved by at least 10 grid cells to avoid artificial effects such as overheating and unphysical flattening of planetary envelopes. In their study of circumplanetary disks, they primarily used a smoothing length of 0.1 Hill radii, though they found that using a smaller smoothing length of 0.025 Hill radii (corresponding to a deeper potential) significantly impacted the temperature structure near the planet and the inner boundary of the CPD. Their results emphasize the importance of carefully choosing and adequately resolving the smoothing length to capture the true physical behavior of gas around forming planets.

\subsection{Angular momentum flux}
\begin{figure*}[hbt!]
% \vspace{-2cm}
    \includegraphics[width=\textwidth]{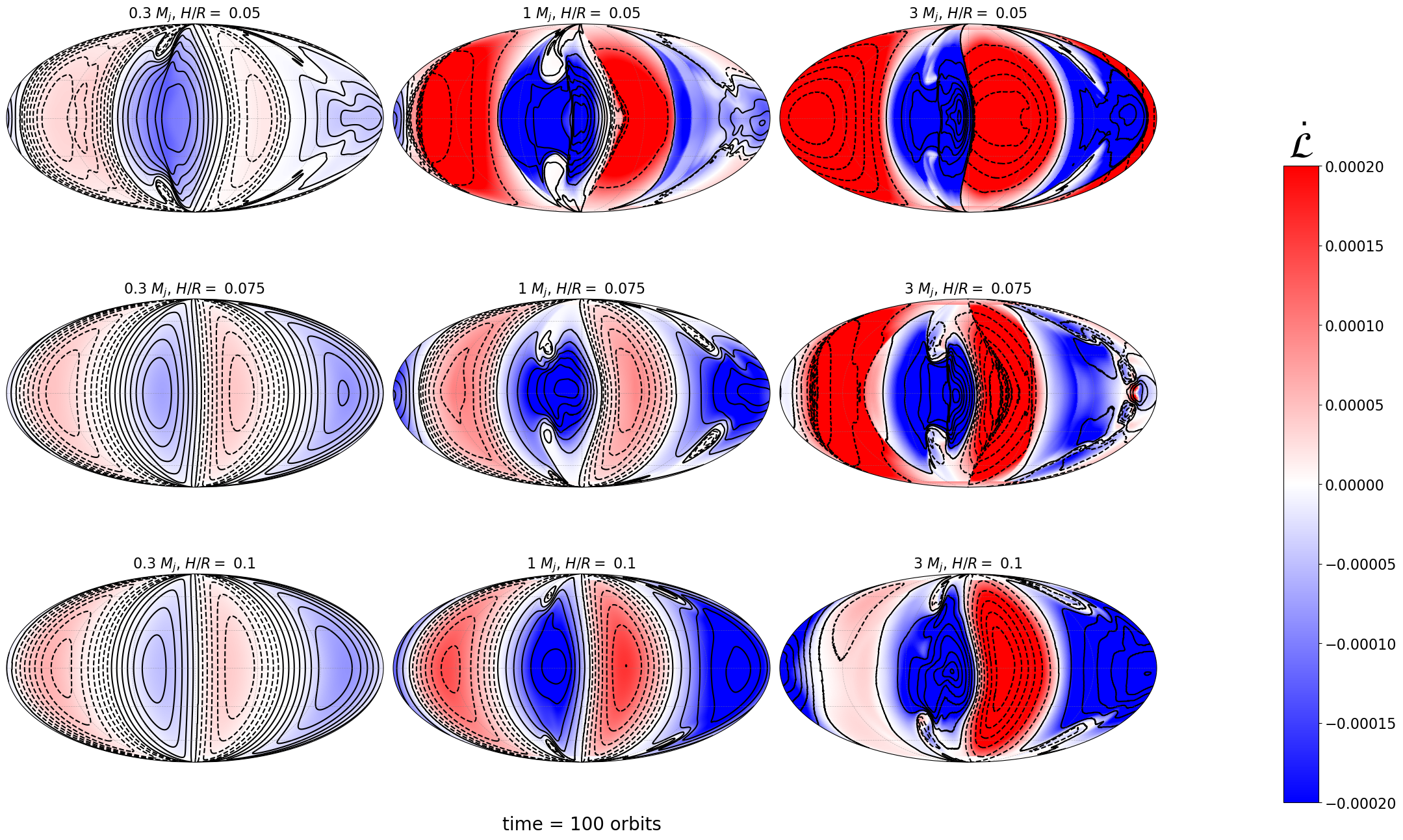}
    % \vspace{-3cm}
        \caption{Angular momentum flux across the Hill sphere for planets of different masses in disks with varying aspect ratios, displayed using Mollweide projections.  To better understand the 3D geometry, consider that the central vertical line in each panel corresponds to the view as if looking at the Hill sphere in the x-z plane, with the planet's orbital motion perpendicular to this plane. Each panel represents a different combination of planet mass and disk aspect ratio, as labeled. The color scale indicates the magnitude and direction of angular momentum flux, with red representing positive (outward) flux and blue representing negative (inward) flux. The black solid contours show the mass flux outflow, while the dashed contours show the inflow. All snapshots are taken at 100 orbital periods.}
        \label{fig:angular-mom}
\end{figure*}

The angular momentum flux across the Hill sphere further illustrates the complex interplay between the protoplanet and its surrounding disk (Figure \ref{fig:angular-mom}). One of the features of the panels in Figure \ref{fig:angular-mom} is the clear asymmetry in the angular momentum flux. We observe distinct positive and negative flux regions. The regions of negative flux closer to the poles represent outflows or recycling of material from the CPD back into the protoplanetary disk. For the lowest mass planets (0.3 Jupiter masses), especially at higher aspect ratios, we see less structured patterns. The development of complex, multi-banded structures in the higher mass cases suggests the formation of intricate flow patterns around the planet, which could be the effects of spiral arms. We see strong, localized regions of positive flux near the poles for the highest mass planets (3 $M_J$) at lower aspect ratios. This is consistent with Figure \ref{fig:vertical-cuts} that suggests that the planet is actively accreting material from high latitudes in these cases.
The variation in flux patterns with disk aspect ratio highlights the importance of disk structure in planet-disk interactions. The more diffuse patterns in thicker disks suggest that planet-disk interactions are spread over a larger volume, potentially leading to weaker but more extended perturbations in the disk.

\begin{figure}[hbt!]
% \vspace{-2cm}
    \includegraphics[width=0.46\textwidth]{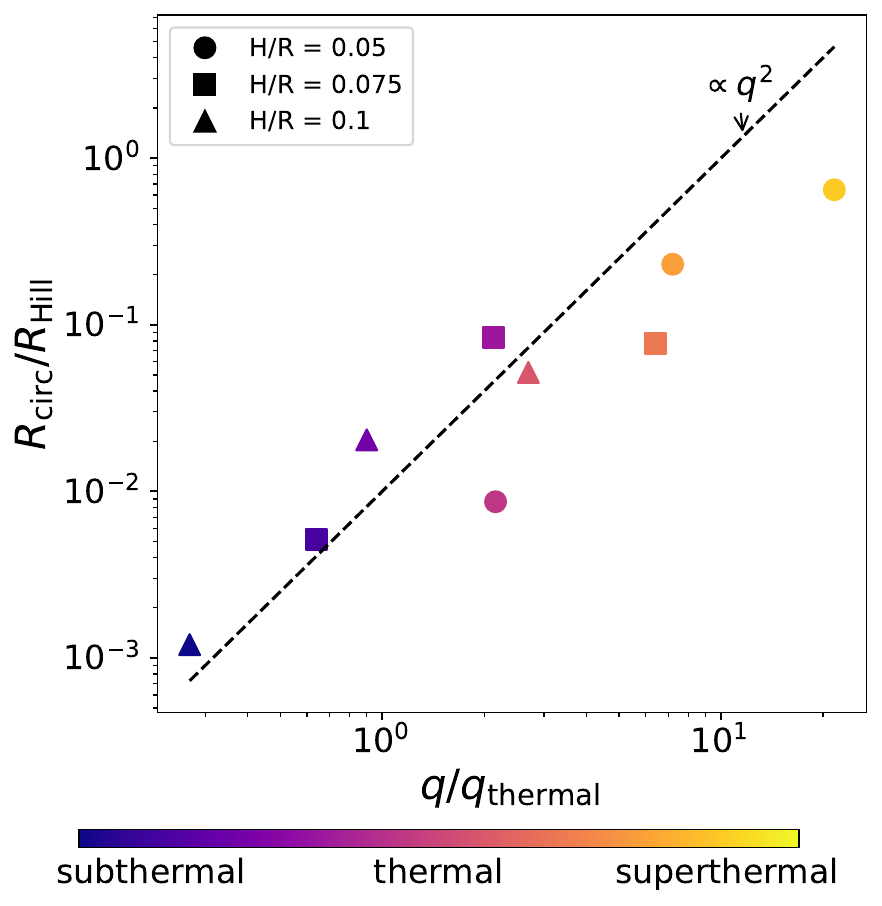}
    % \vspace{-3cm}
        \caption{Relationship between the circularization radius ($R_{\rm circ}$) normalized by the Hill radius ($R_{\rm Hill}$) and the thermal mass ratio ($q/q_{\rm thermal}$) for forming planets. The x-axis shows $q/q_{\rm thermal}$ on a logarithmic scale, spanning from subthermal ($q/q_{\rm thermal}$) to superthermal ($q/q_{\rm thermal}$) regimes. The y-axis represents $R_{\rm circ}/R_{\rm Hill}$. The plot demonstrates how the relative size of the CPD (represented by $R_{\rm circ}$) changes with planet mass relative to the disk's thermal mass, highlighting the complex interplay between planetary gravity and disk thermal properties in shaping CPD structure.}
        \label{fig:qthermal-rcirc}
\end{figure}

Figure \ref{fig:qthermal-rcirc} shows the relationship between the circularization radius (Equation \ref{eq:rcirc}) normalized by the Hill radius and the thermal mass ratio. The circularization radius represents the radius at which infalling material from the Hill sphere would settle into circular orbits around the planet if angular momentum was conserved. Material falling inside this radius is likely to become part of the CPD, while material outside may be more easily lost back to the protoplanetary disk. There's a power-law trend in the $R_{\rm circ}/R_{\rm Hill}$ ratio across different planet masses and disk aspect ratios. Higher aspect ratios for a given planet mass generally lead to smaller $R_{\rm circ}/R_{\rm Hill}$ ratios. This indicates that the material needs to get closer to the planet in thicker disks before it can form a stable CPD, possibly due to increased pressure support in the background disk.
More massive planets tend to have larger $R_{\rm circ}/R_{\rm Hill}$ ratios. This trend suggests that a larger fraction of the Hill sphere contributes to the formation of the CPD for more massive planets. This is an interesting finding as it implies that more massive planets might be more effective at capturing and retaining material from their surrounding environment to form extensive circumplanetary disks.

\section{Discussion} 
\label{sec:discussion}
\subsection{Global disk morphology}
As discussed in \cite{Miranda2020}, the concept of thermal mass provides important insights into how planets of different masses interact with the protoplanetary disk and potentially affect the CPD formation. 

For sub-thermal mass planets ($q < q_{\rm thermal}$), the density waves excited in the disk remain largely linear and oscillatory. These waves transport angular momentum but do not significantly dissipate energy or strongly perturb the disk structure. As a result, gap opening is limited and the disk maintains a more continuous profile around the planet's orbit. This can be seen in our Figure \ref{fig:gap-profiles}, especially evident in the $H/R = 0.1$ case, where the gaps are not as deep as in other cases. In this regime, CPD formation might occur in a relatively unperturbed local disk environment. 

As planet mass approaches and exceeds the thermal mass ($q \leq q_{\rm thermal}$), the excited density waves become nonlinear, steepening into shocks as they propagate. This nonlinear evolution leads to more effective angular momentum transport and energy dissipation in the disk. Therefore, more pronounced gaps open around the planet's orbit, potentially creating pressure maxima that trap dust and alter the local disk structure. The pronounced gap edges in super-thermal cases represent pressure maxima that enhance the concentration of solids near the planet, potentially providing a reservoir of material for CPD formation. These pressure maxima regions efficiently trap drifting pebbles, leading to significant solid enhancement. This concentration of material facilitates the formation of planetesimals through streaming instability and affects the global distribution of dust in the disk.
The efficiency of this trapping mechanism depends on several factors, including the planet's mass and disk properties. For planets above the pebble isolation mass, these pressure maxima can act as nearly impenetrable barriers to inward drifting pebbles, potentially leading to the formation of dust-depleted inner cavities resembling transition disks (see Figure \ref{fig:vertical-cuts}). The material concentrated at these gap edges could serve multiple purposes in planetary system evolution. Besides providing raw material for satellite formation, it could also contribute to the late-stage growth of the planet itself \citep{Eriksson2022, Zhu2012}. Furthermore, the enhanced solid density in these regions could lead to the formation of additional planetesimals, potentially seeding a new generation of planetary bodies or contributing to debris disk formation. The transition from shallow to deep gaps as planet mass increases likely reflects increased nonlinear dissipation of density waves near the planet, which could drive local angular momentum transport and influence material accretion onto the CPD. These varying gap depths and profiles across different $H/R$ values in the simulations highlight the importance of disk thermodynamics in shaping the environment where CPDs form \citep{Miranda2020, Krapp2024}. 

In Figure \ref{fig:midplane-cuts}, we can see the development of an eccentric gap in the case of larger planets, e.g., the case of a 3 Jupiter-mass planet with $H/R=0.075$. This observation is consistent with \cite{Teyssandier2017}, who identified a threshold for eccentricity growth in disks with gap-opening planets. They found that eccentricity growth occurs for planet-to-star mass ratios exceeding $q \approx 3 \times 10^{-3}$, corresponding to about 3 Jupiter masses for a solar-mass star.
However, it's important to note that our model with an eccentric gap has a different aspect ratio ($H/R=0.075$) compared to their study ($H/R=0.05$). This difference could affect the precise threshold for eccentricity growth, as the aspect ratio influences the strength of pressure effects and the width of resonances in the disk. 
The eccentric gaps formed by massive planets create a dynamic environment that likely influences the flow of material onto the CPD. As the gap becomes eccentric, the distance between the planet and the gap edge varies azimuthally because of precession. This variation could lead to periodic changes in the mass flow towards the planet, potentially causing accretion variability \citep{Teyssandier2017}.
In our simulations, we measured mass flux onto the CPD throughout 100 orbits. However, we did not observe significant variability in this flux that could be attributed to gap eccentricity. This is likely because 100 orbits are insufficient for substantial eccentricity to develop. \cite{Teyssandier2017} showed for similar mass ratios, the eccentric instability takes between 100-1000 orbits to fully grow.

Moreover, as spiral arms launched by the planet interact with the eccentric gap edge, they may create localized regions of enhanced density and pressure (see, e.g., Figure \ref{fig:spirals}). When these high-density regions align with the planet's position, they could lead to episodic increases in accretion onto the CPD \citep{Kley2006, Dunhill2013}. These regions of higher eccentricity could lead to more complex orbital dynamics of the accreting material. As a result, the angular momentum of the inflowing gas might vary significantly depending on its point of origin along the eccentric gap edge. This variation could impact the efficiency of angular momentum transport within the CPD \citep{Lubow2006, Machida2010, Ragusa2018}.

\subsection{CPD Inflow Geometry}
Our simulations reveal a clear dependence of CPD and envelope structures on both planetary mass and disk aspect ratio. A key finding is the transition from envelope-dominated to disk-dominated structures as planetary mass increases. Lower-mass planets, particularly those below the thermal mass ($q < q_{\rm thermal}$), tend to be surrounded by quasi-spherical envelopes rather than well-defined disks. This envelope morphology likely results from the thermal energy of the gas counterbalancing the planet's gravitational influence, preventing the formation of a flattened disk structure. This interpretation aligns with \cite{Krapp2024}, which emphasizes the significant role of cooling time in shaping disk formation. They suggest that the time to dissipate thermal energy is critical for the formation of a disk, particularly in lower-mass planets where longer cooling times may favor the maintenance of envelope dominance rather than transitioning to disk-dominated configurations.

Moreover, recent work by \cite{Lega2024} finds that gas accretion rates around Jupiter-mass planets scale with $\sqrt{\Sigma}\nu$, rather than scaling directly with surface density and viscosity. Their work demonstrates that efficient cooling is critical for CPD formation, with lower-mass and low-viscosity disks being more favorable for developing rotationally-supported CPD structures. This complements our results about the importance of disk properties, suggesting that both geometric factors ($H/R$ ratio) and thermodynamic properties (cooling efficiency) play crucial roles in determining CPD formation and structure. Particularly relevant to our study, \cite{Lega2024} find that when using a fully radiative equation of state with constant dust-to-gas ratio, lower-mass and lower-viscosity protoplanetary disks favor the formation of rotationally supported CPDs.

In contrast, higher-mass planets, especially those above the thermal mass, develop clear disk-like morphologies. These CPDs are characterized by flattened structures with increased density towards the midplane and often feature prominent spiral patterns. The emergence of these patterns suggests the presence of mechanisms for angular momentum transport within the disk, which could play a role in the disk's evolution and potential for satellite formation.

In our simulations, Saturn-mass planets (0.3 $M_J$) generally do not form well-defined CPDs, especially in disks with higher aspect ratios. Instead, these planets are surrounded by more spherical, envelope-like structures. If Saturn formed in a region of the protoplanetary disk with a relatively high aspect ratio ($H/R \geq 0.075$), our results suggest that it may not have developed a substantial circumplanetary disk during its early stages of growth. In such a scenario, the formation of Saturn's regular satellites might have been delayed until the planet migrated to a region of the disk with a lower aspect ratio or until the overall disk cooled and thinned over time.

Alternatively, if Saturn formed in a cooler, thinner region of the protoplanetary disk ($H/R \sim 0.05$), our simulations indicate that a more disk-like structure could have developed around the planet. For instance, the Grand Tack model \citep{Walsh2011} suggests that Saturn initially formed around 8-9 AU before migrating inward and then outward. In contrast, the Nice model and its variants \citep{Desch2018} typically place Saturn's formation further out, around 11-12 AU. Both scenarios could potentially provide the cooler, thinner disk conditions ($H/R \sim 0.05$) that our simulations suggest are favorable for the formation of a disk-like structure around Saturn. We note, however, that even for lower aspect ratio protoplanetary disks, our estimates of the minimum planet mass for CPD formation are typically well above the pebble isolation mass \citep{Lambrechts2014,Bitsch2018}. The ability of the protoplanetary disk to deliver all but the smallest solid particles may therefore be limited, and the full implications for regular satellite formation require global simulations with dust as well as gas physics.

The disk aspect ratio ($H/R$) significantly influences CPD morphology. Higher aspect ratios generally lead to more vertically extended structures and less sharply defined disks, which highlights the importance of the background disk's thermal structure in shaping the circumplanetary environment. Lower aspect ratios, on the other hand, result in thinner, more clearly defined CPDs, particularly for higher-mass planets.

Furthermore, our results indicate that CPD formation is not a binary outcome but rather a continuum, with the degree of "diskiness" increasing smoothly with planetary mass and decreasing disk aspect ratio (see Figure \ref{fig:keplerian}). This gradual transition challenges simplistic models of CPD formation and emphasizes the need for nuanced approaches in studying these systems.

The complex interplay between planetary mass, disk properties, and CPD morphology underscores the importance of considering a wide range of parameters in giant planets and satellite formation models. Our results suggest that the conditions for satellite formation may exhibit a greater diversity than previously thought, with potential implications for the diversity of exomoon systems we might expect to observe in the future.

We also analyze a column density profile in Appendix \ref{appendix} \citep{Taylor2024}.

\subsection{Mass and angular momentum flux}
The mass and angular momentum flux patterns observed in our Saturn-mass simulations also provide clues about the planet's potential migration history. The relatively uniform mass inflow across the Hill sphere for these lower-mass planets suggests that Saturn may have experienced smoother, more continuous accretion during its formation compared to higher-mass planets like Jupiter. This could have implications for the timescale and efficiency of Saturn's growth. 
% Based on our simulations, the accretion timescale for Saturn-mass planets appears to be on the order of $10^6$ to $10^7$ years ($t_{acc} \sim \frac{0.3 M_J}{10^{-4} M_J / \rm orbit} \times 100 \rm orbits \times 29.5 \rm years$). 
This is consistent with recent models of giant planet formation that suggest rapid growth once the core reaches a critical mass \citep{Lissauer2009}.

If Saturn underwent significant migration during its formation, our results suggest that its ability to accrete and retain material may have varied depending on its location within the protoplanetary disk. Current models of solar system formation, such as the Nice model and its variants, suggest that Saturn may have formed closer to the Sun than its current orbit, possibly in the region between 5 and 8 AU \citep{Tsiganis2005, Morbidelli2007, Walsh2011}. Our simulations indicate that in this inner region, where the aspect ratio of the protoplanetary disk was likely higher due to stronger stellar irradiation, Saturn would have been surrounded by a more envelope-like structure. As Saturn migrated outward to its current position at about 9.5 AU, it would have encountered regions of the disk with progressively lower aspect ratios. This migration from a region with a higher aspect ratio to one with a lower aspect ratio could have led to a transition from an envelope-dominated to a more disk-like circumplanetary structure, potentially triggering episodes of enhanced satellite formation. Such a scenario could explain the diverse compositions and characteristics observed in Saturn's current satellite system, with different moons forming under varying conditions as Saturn migrated \citep{Crida2006, Canup2006}.

A key metric in understanding the extent of CPD formation is the circularization radius ($R_{\rm circ}$), which represents the radius at which infalling material from the Hill sphere would settle into circular orbits around the planet. Our results show a clear trend of increasing $R_{\rm circ}/R_{\rm Hill}$ ratio with increasing planet mass relative to the thermal mass. A larger ratio suggests that more massive planets can support circular orbits over a greater portion of their Hill spheres, potentially leading to more extensive CPDs.

Interestingly, we find that the $R_{\rm circ}/R_{\rm Hill}$ ratio is generally smaller for higher disk aspect ratios at a given planet mass. This implies that in thicker disks, the material needs to get closer to the planet before it can form a stable CPD, possibly due to increased pressure support in the background disk. The relationship between $R_{\rm circ}/R_{\rm Hill}$l and $q/q_{\rm thermal}$ is consistent with a power-law of $\approx (q/q_{\rm thermal})^2$ trend.

As the planet's mass increases, there's a general upward trend, indicating that more massive planets can accumulate a larger fraction of the available material within their Hill sphere into their CPD. This trend is particularly steep in the subthermal regime, suggesting that as young planets grow, they become increasingly effective at capturing and retaining mass from the surrounding protoplanetary disk.

It also looks like thinner disks might lead to reduced mass flux into the CPD. In a geometrical sense, a thinner disk provides a less vertical extent of material for the planet to accrete and may have steeper pressure gradients, potentially affecting the flow of material toward the planet. The interplay between the Hill radius and disk scale height becomes important when the Hill radius exceeds the scale height, as the planet's gravitational influence extends beyond the disk's vertical extent, potentially altering accretion dynamics.

\section{Conclusions}
\label{sec:conclusions}
Our three-dimensional hydrodynamic simulations of planet-disk interactions provide clues into the formation and structure of circumplanetary disks (CPDs) across a range of planetary masses and disk properties. Key findings include:
\begin{itemize}
    \item The transition from envelope-dominated to disk-dominated structures around forming planets occurs along a continuum that is primarily governed by the planet's mass relative to the disk's thermal mass ($q/q_{\rm thermal}$) and the disk's aspect ratio ($H/R$). The local disk conditions ($H/R$) are important in determining whether embedded planets can host a rotationally supported CPD. 
    \item More massive planets, particularly those above the thermal mass, tend to form well-defined, flattened CPDs with prominent spiral patterns. In contrast, lower-mass planets (below thermal mass) are often surrounded by quasi-spherical envelopes rather than disks.
    \item Higher disk aspect ratios lead to more vertically extended structures and less sharply defined CPDs, while lower aspect ratios result in thinner, more clearly defined disks, especially for higher-mass planets.
    \item The ratio of circularization radius to Hill radius ($R_{\rm circ}/R_{\rm Hill}$) increases with planet mass relative to thermal mass, suggesting that more massive planets are more effective at capturing and retaining material to form extensive CPDs.
    
\end{itemize}

Our results have implications for our understanding of giant planet and satellite formation:
\begin{itemize}
    \item The diversity in CPD structures suggests that satellite formation conditions may vary significantly as a function of planetary orbital radius (through the scaling of $H/R$ with distance from the star), as well as mass, leading to a diversity of satellite systems.
    \item The detectability and characteristics of CPDs around exoplanets will vary significantly based on the planet's mass and the properties of the protoplanetary disk.
    \item The most favorable circumstance for forming detectable CPDs around planets at large orbital separation requires a massive planet, orbiting in a shadowed region where the circumplanetary gas can cool (possibly to a temperature below that of the nearby protoplanetary disk). The low abundance of detected CPDs may be attributed to the combined effect of a low frequency of super-thermal mass protoplanets, orbiting at distances sufficiently far from the star to admit direct imaging detections.
    \item Jupiter likely had a substantial CPD during its formation, whereas Saturn's capacity to form a substantial CPD and, consequently, its regular satellites may have been highly dependent on its location and migration history within the protoplanetary disk. 
\end{itemize}

Potentially fruitful avenues for future work include the incorporation of more realistic thermodynamics \citep{Krapp2024}, exploring the effects of planetary migration, and understanding how to use simulations to inform models of the long-term evolution of CPDs and their potential for satellite formation. 

\acknowledgments
SS thanks Wenrui Xu and Nicolas Cimerman for valuable conversations. SS also acknowledges the Texas Advanced Computing Center of the University of Texas at Austin and the Computation for the Endless Frontier (NSF Award 1818253). PJA acknowledges support from NASA TCAN award 80NSSC19K0639, and from award 644616 from the Simons Foundation. RL acknowledges support from the Heising-Simons Foundation 51 Pegasi b Fellowship.

\bibliographystyle{aasjournal}

\begin{thebibliography}{99}

\bibitem[Speedie et al.(2024)]{Speedie2024}
Speedie, J., Dong, R., Hall, C., et al. 2024, Nature, 633, 58

\bibitem[Martin et al.(2023)]{Martin2023}
Martin, R. G., Armitage, P. J., Lubow, S. H., \& Price, D. J. 2023, ApJ, 953, 2

\bibitem[Lubow \& Martin(2013)]{Lubow2013}
Lubow, S. H., \& Martin, R. G. 2013, MNRAS, 428, 2668

\bibitem[Quillen \& Trilling(1998)]{Quillen1998}
Quillen, A. C., \& Trilling, D. E. 1998, ApJ, 508, 707

\bibitem[Lubow et al.(1999)]{Lubow1999}
Lubow, S. H., Seibert, M., \& Artymowicz, P. 1999, ApJ, 526, 1001

\bibitem[Ayliffe \& Bate(2009)]{Ayliffe2009}
Ayliffe, B. A., \& Bate, M. R. 2009, MNRAS, 397, 657

\bibitem[Morbidelli et al.(2014)]{Morbidelli2014}
Morbidelli, A., Szulágyi, J., Crida, A., et al. 2014, Icarus, 232, 266

\bibitem[D'Angelo et al.(2003)]{Dangelo2023b}
D'Angelo, G., Kley, W., \& Henning, T. 2003, ApJ, 586, 540

\bibitem[Bae et al.(2022)]{Bae2022}
Bae, J., Teague, R., Andrews, S. M., et al. 2022, ApJL, 934, L20

\bibitem[Benisty et al.(2021)]{Benisty2021}
Benisty, M., Bae, J., Facchini, S., et al. 2021, ApJL, 916, L2

\bibitem[Andrews(2020)]{Andrews2020}
Andrews, S. M. 2020, ARA\&A, 58, 483

\bibitem[Andrews et al.(2021)]{Andrews2021}
Andrews, S. M., Elder, W., Zhang, S., et al. 2021, ApJ, 916, 51

\bibitem[Perez et al.(2015)]{Perez2015}
Perez, S., Dunhill, A., Casassus, S., et al. 2015, ApJL, 811, L5

\bibitem[Szulágyi et al.(2019)]{Szulagyi2019}
Szulágyi, J., Dullemond, C. P., Pohl, A., \& Quanz, S. P. 2019, MNRAS, 487, 1248

\bibitem[Eisner(2015)]{Eisner2015}
Eisner, J. A. 2015, ApJL, 803, L4

\bibitem[Zhu(2015)]{Zhu2015}
Zhu, Z. 2015, ApJ, 799, 16

\bibitem[Martin \& Lubow(2011)]{Martin2011}
Martin, R. G., \& Lubow, S. H. 2011, MNRAS, 413, 1447

\bibitem[Pollack et al.(1996)]{Pollack1996}
Pollack, J. B., Hubickyj, O., Bodenheimer, P., et al. 1996, Icarus, 124, 62

\bibitem[Batygin \& Morbidelli(2020)]{Batygin2020}
Batygin, K., \& Morbidelli, A. 2020, ApJ, 894, 143

\bibitem[Fung et al.(2019)]{Fung2019}
Fung, J., Zhu, Z., \& Chiang, E. 2019, ApJ, 887, 152

\bibitem[Teague et al.(2019)]{Teague2019}
Teague, R., Bae, J., \& Bergin, E. A. 2019, Nature, 574, 378

\bibitem[Szulágyi et al.(2019)]{Szul2019}
Szulágyi, J., Dullemond, C. P., Pohl, A., \& Quanz, S. P. 2019, MNRAS, 487, 1248

\bibitem[Isella et al.(2019)]{Isella2019}
Isella, A., Benisty, M., Teague, R., et al. 2019, ApJL, 879, L25

\bibitem[Ward-Duong et al.(2018)]{Ward2018}
Ward-Duong, K., Patience, J., Bulger, J., et al. 2018, AJ, 155, 54

\bibitem[Szul\'{a}gyi(2017)]{Szul2017b}
Szul\'{a}gyi, J. 2017, ApJ, 842, 103

\bibitem[Fujii et al.(2017)]{Fujii2017}
Fujii, Y. I., Kobayashi, H., Takahashi, S. Z., \& Gressel, O. 2017, AJ, 153, 194

\bibitem[Szul\'{a}gyi et al.(2017)]{Szul2017a}
Szulágyi, J., Mayer, L., \& Quinn, T. 2017, MNRAS, 464, 3158

\bibitem[Szul\'{a}gyi et al.(2016)]{Szul2016}
Szulágyi, J., Masset, F., Lega, E., et al. 2016, MNRAS, 460, 2853

\bibitem[Zhu \& Baruteau(2016)]{Zhu2016}
Zhu, Z., \& Baruteau, C. 2016, MNRAS, 458, 3918

\bibitem[Fujii et al.(2014)]{Fujii2014}
Fujii, Y. I., Okuzumi, S., Tanigawa, T., \& Inutsuka, S. 2014, ApJ, 785, 101

\bibitem[Szulágyi et al.(2014)]{Szul2014}
Szulágyi, J., Morbidelli, A., Crida, A., \& Masset, F. 2014, ApJ, 782, 65

\bibitem[Gressel et al.(2013)]{Gressel2013}
Gressel, O., Nelson, R. P., Turner, N. J., \& Ziegler, U. 2013, ApJ, 779, 59

\bibitem[Fujii et al.(2011)]{Fujii2011}
Fujii, Y. I., Okuzumi, S., \& Inutsuka, S. 2011, ApJ, 743, 53

\bibitem[Ward \& Canup(2010)]{Ward2010}
Ward, W. R., \& Canup, R. M. 2010, AJ, 140, 1168

\bibitem[Machida et al.(2008)]{Machida2008}
Machida, M. N., Kokubo, E., Inutsuka, S., \& Matsumoto, T. 2008, ApJ, 685, 1220

\bibitem[D'Angelo et al.(2003)]{Dangelo2003}
D'Angelo, G., Henning, T., \& Kley, W. 2003, ApJ, 599, 548

\bibitem[Fung et al.(2017)]{Fung_2017}
Fung, J., Masset, F., Lega, E., \& Velasco, D. 2017, AJ, 153, 124

\bibitem[Lunine \& Stevenson(1982)]{Lunine1982}
Lunine, J. I., \& Stevenson, D. J. 1982, Icarus, 52, 14

\bibitem[Goodman \& Rafikov(2001)]{Goodman2001}
Goodman, J., \& Rafikov, R. R. 2001, ApJ, 552, 793

\bibitem[Canup \& Ward(2006)]{Canup2006}
Canup, R. M., \& Ward, W. R. 2006, Nature, 441, 834

\bibitem[Heller \& Pudritz(2015a)]{Heller2015a}
Heller, R., \& Pudritz, R. 2015, ApJ, 806, 181

\bibitem[Heller \& Pudritz(2015b)]{Heller2015b}
Heller, R., \& Pudritz, R. 2015, A\&A, 578, A19

\bibitem[Shibaike et al.(2017)]{Shibaike2017}
Shibaike, Y., Okuzumi, S., Sasaki, T., \& Ida, S. 2017, ApJ, 846, 81

\bibitem[Ronnet et al.(2018)]{Ronnet2018}
Ronnet, T., Mousis, O., Vernazza, P., et al. 2018, AJ, 155, 224

\bibitem[Krapp et al.(2024)]{Krapp2024}
Krapp, L., Kratter, K. M., Youdin, A. N., et al. 2024, arXiv:2402.14638

\bibitem[Canup \& Ward(2009)]{Canup2009}
Canup, R. M., \& Ward, W. R. 2009, in Europa, ed. R. T. Pappalardo, W. B. McKinnon, \& K. K. Khurana, 59

\bibitem[Estrada et al.(2009)]{Estrada2009}
Estrada, P. R., Mosqueira, I., Lissauer, J. J., et al. 2009, in Europa, ed. R. T. Pappalardo, W. B. McKinnon, \& K. K. Khurana, 27

\bibitem[Canup \& Ward(2002)]{Canup2002}
Canup, R. M., \& Ward, W. R. 2002, AJ, 124, 3404

\bibitem[Mosqueira \& Estrada(2003a)]{Mosqueira2003a}
Mosqueira, I., \& Estrada, P. R. 2003, Icarus, 163, 232

\bibitem[Mosqueira \& Estrada(2003b)]{Mosqueira2003b}
Mosqueira, I., \& Estrada, P. R. 2003, Icarus, 163, 198

\bibitem[Miguel \& Ida(2016)]{Miguel2016}
Miguel, Y., \& Ida, S. 2016, Icarus, 266, 1

\bibitem[Maeda et al.(2022)]{Maeda2022}
Maeda, N., Ohtsuki, K., Tanigawa, T., et al. 2022, ApJ, 935, 56

\bibitem[Maeda et al.(2024)]{Maeda2024}
Maeda, N., Ohtsuki, K., Suetsugu, R., et al. 2024, arXiv:2404.11857

\bibitem[Raymond \& Cossou(2014)]{Raymond2014}
Raymond, S. N., \& Cossou, C. 2014, MNRAS, 440, L11

\bibitem[Chambers(2009)]{Chambers2009}
Chambers, J. E. 2009, ApJ, 705, 1206

\bibitem[Stone et al.(2020)]{Stone2020}
Stone, J. M., Tomida, K., White, C. J., \& Felker, K. G. 2020, ApJS, 249, 4

\bibitem[Kley(1999)]{Kley1999}
Kley, W. 1999, MNRAS, 303, 696

\bibitem[Tanigawa et al.(2012)]{Tanigawa2012}
Tanigawa, T., Ohtsuki, K., \& Machida, M. N. 2012, ApJ, 747, 47

\bibitem[Shakura \& Sunyaev(1973)]{Shakura1973}
Shakura, N. I., \& Sunyaev, R. A. 1973, A\&A, 24, 337

\bibitem[Crida et al.(2006)]{Crida2006}
Crida, A., Morbidelli, A., \& Masset, F. 2006, Icarus, 181, 587

\bibitem[Fung et al.(2014)]{Fung2014}
Fung, J., Shi, J.-M., \& Chiang, E. 2014, ApJ, 782, 88

\bibitem[Taylor \& Adams(2024)]{Taylor2024}
Taylor, A. G., \& Adams, F. C. 2024, Icarus, 415, 116044

\bibitem[Miranda \& Rafikov(2020)]{Miranda2020}
Miranda, R., \& Rafikov, R. R. 2020, ApJ, 904, 121

\bibitem[Teyssandier \& Ogilvie(2017)]{Teyssandier2017}
Teyssandier, J., \& Ogilvie, G. I. 2017, MNRAS, 467, 4577

\bibitem[Zhou et al.(2022)]{Zhou2022}
Zhou, Y., Sanghi, A., Bowler, B. P., et al. 2022, ApJL, 934, L13

\bibitem[Currie et al.(2022)]{Currie2022}
Currie, T., Lawson, K., Schneider, G., et al. 2022, Nature Astronomy, 6, 751

\bibitem[Lissauer et al.(2009)]{Lissauer2009}
Lissauer, J. J., Hubickyj, O., D'Angelo, G., \& Bodenheimer, P. 2009, Icarus, 199, 338

\bibitem[Tsiganis et al.(2005)]{Tsiganis2005}
Tsiganis, K., Gomes, R., Morbidelli, A., \& Levison, H. F. 2005, Nature, 435, 459

\bibitem[Morbidelli et al.(2007)]{Morbidelli2007}
Morbidelli, A., Tsiganis, K., Crida, A., et al. 2007, AJ, 134, 1790

\bibitem[Walsh et al.(2011)]{Walsh2011}
Walsh, K. J., Morbidelli, A., Raymond, S. N., et al. 2011, Nature, 475, 206

\bibitem[Zhu et al.(2012)]{Zhu2012}
Zhu, Z., Nelson, R. P., Dong, R., et al. 2012, ApJ, 755, 6

\bibitem[Eriksson et al.(2020)]{Eriksson2022}
Eriksson, L. E. J., Johansen, A., \& Liu, B. 2020, A\&A, 635, A110

\bibitem[Kley \& Dirksen(2006)]{Kley2006}
Kley, W., \& Dirksen, G. 2006, A\&A, 447, 369

\bibitem[Dunhill et al.(2013)]{Dunhill2013}
Dunhill, A. C., Alexander, R. D., \& Armitage, P. J. 2013, MNRAS, 428, 3072

\bibitem[Lubow \& D'Angelo(2006)]{Lubow2006}
Lubow, S. H., \& D'Angelo, G. 2006, ApJ, 641, 526

\bibitem[Ragusa et al.(2018)]{Ragusa2018}
Ragusa, E., Rosotti, G., Teyssandier, J., et al. 2018, MNRAS, 474, 4460

\bibitem[Machida et al.(2010)]{Machida2010}
Machida, M. N., Kokubo, E., Inutsuka, S., \& Matsumoto, T. 2010, MNRAS, 405, 1227

\bibitem[Desch et al.(2018)]{Desch2018}
Desch, S. J., Kalyaan, A., \& Alexander, C. M. O'D. 2018, ApJS, 238, 11

\bibitem[Zhou et al.(2023)]{Zhou2023}
Zhou, Y., Bowler, B. P., Yang, H., et al. 2023, AJ, 166, 220

\bibitem[Schulik et al.(2020)]{Schulik2020}
Schulik, M., Johansen, A., Bitsch, B., Lega, E., \& Lambrechts, M. 2020, A\&A, 642, A187

\bibitem[Lega et al.(2024)]{Lega2024}
Lega, E., Benisty, M., Cridland, A., Morbidelli, A., Schulik, M., \& Lambrechts, M. 2024, A\&A, 690, A183

\bibitem[Bitsch et al.(2018)]{Bitsch2018}
Bitsch, B., Morbidelli, A., Johansen, A., Lega, E., Lambrechts, M, \& Crida, A. 2018, A\&A, 612, A30

\bibitem[Lambrechts et al.(2014)]{Lambrechts2014}
Lambrechts, M. \& Johansen, A. 2014, A\&A, 572, A107

\end{thebibliography}

\end{CJK*}

\appendix
\section{Comparison to an analytical inflow solution}
\label{appendix}
\begin{figure*}[hbt!]
% \vspace{-2cm}
    \center
    \includegraphics[width=0.9\textwidth]{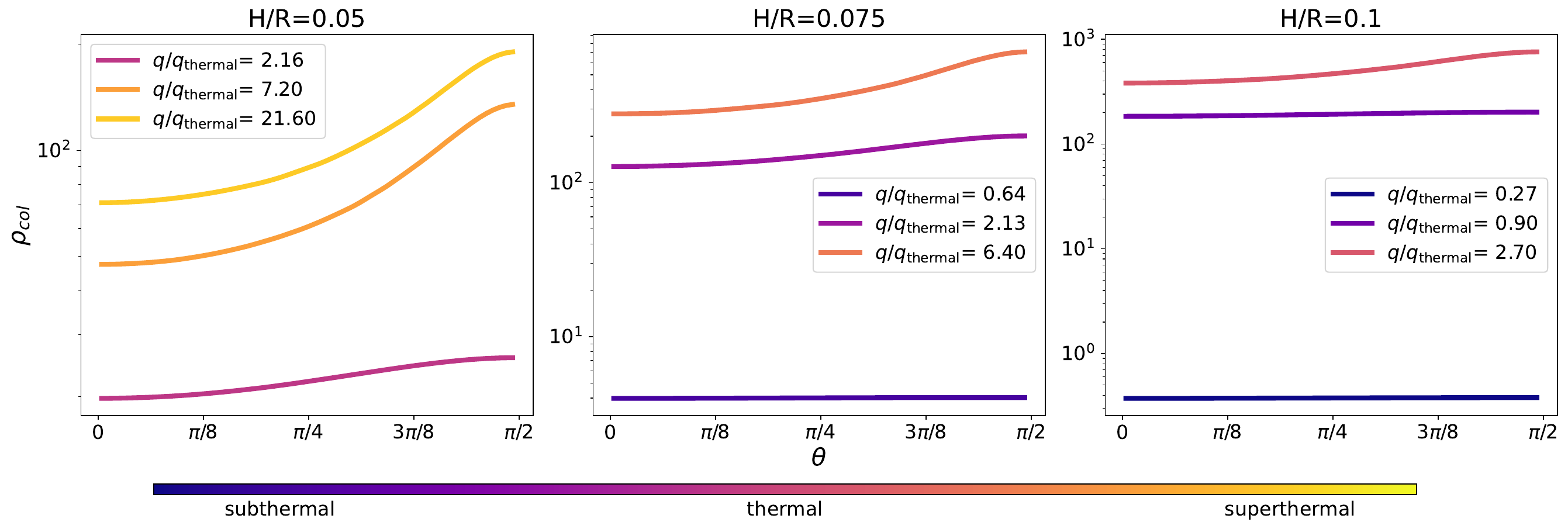}
    % \vspace{-3cm}
        \caption{Column density profiles as a function of polar angle for planets with different masses in disks of varying aspect ratios. Each panel corresponds to a specific disk aspect ratio (H/R = 0.05, 0.075, and 0.1, from top to bottom). Within each panel, different lines represent planets of varying masses, characterized by their thermal mass ratio ($q/q_{\rm thermal}$). The x-axis shows the polar angle from 0 to $\pi/2$ radians, and the y-axis displays the column density on a logarithmic scale. Planets are categorized as subthermal, thermal, or superthermal based on their $q/q_{\rm thermal}$ value. Note the general trend of increasing column density towards the disk midplane ($\pi/2$) and the dependence of the profile shape on both planet mass and disk aspect ratio.}
        \label{fig:col-density}
\end{figure*}

\begin{figure}[hbt!]
% \vspace{-2cm}
    \center
    \includegraphics[width=0.46\textwidth]{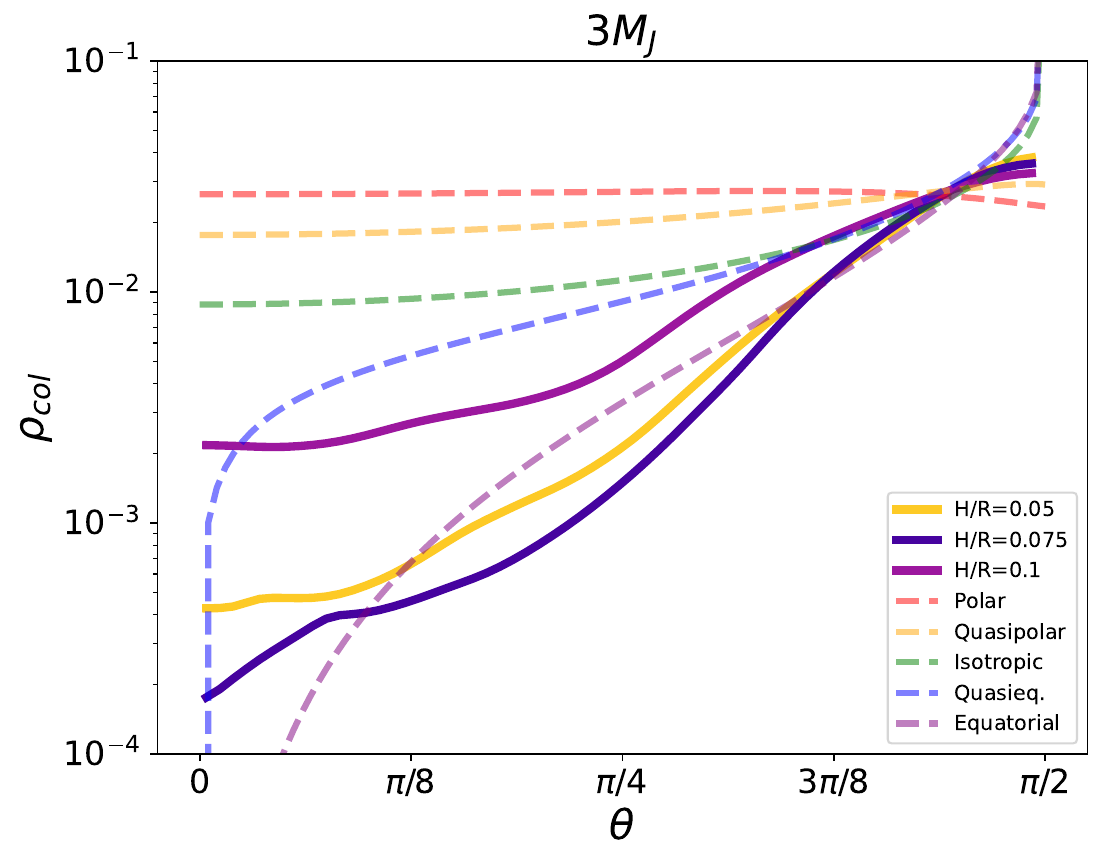}
    % \vspace{-3cm}
        \caption{The same as Figure \ref{fig:col-density}, but only for the 3 Jupiter-mass case and neglecting the pressure support for a better comparison with \cite{Taylor2024}.}
        \label{fig:col-density-3mj}
\end{figure}

Our analysis of column density profiles (Figures \ref{fig:col-density} and \ref{fig:col-density-3mj}) provides further insights into the structure of these circumplanetary environments. We observe a consistent pattern across different planetary masses and disk aspect ratios: a relatively uniform column density across most polar angles, with a sharp increase as the disk mid-plane approaches. This pattern most closely resembles the ``isotropic'' inflow case described in recent analytical models, such as those presented by  \cite{Taylor2024} (Figure 5 in their paper). 

The consistency of this pattern across different planet masses and disk aspect ratios is noteworthy, suggesting that the overall structure of the CPD is robust to changes in these parameters, at least in terms of its angular distribution of material. More massive planets (3 Jupiter masses) generally show higher column densities across all angles, which is expected due to their stronger gravitational influence. The disk aspect ratio ($H/R$) affects the steepness of the increase near the disk plane, with lower aspect ratios (thinner disks) showing a sharper rise, possibly due to material being more concentrated in the disk mid-plane. 

A key difference with the analytical model, however, lies in the role of thermal energy. In our simulations, the thermal energy of the gas plays a significant role in shaping the inflow patterns, particularly for lower-mass planets and in disks with higher aspect ratios. This thermal component leads to a more gradual density gradient and a more uniform inflow pattern than might be expected from purely gravitational considerations.
The thermal energy's influence is especially evident in cases where the planet's mass is below or near the thermal mass ($q \lesssim q_{\rm thermal}$). In these scenarios, the thermal energy of the gas can substantially counteract the planet's gravitational influence, resulting in a more diffuse and less vertically stratified inflow structure. This contrasts with the sharper density gradients and more focused inflow patterns that might be predicted by models that do not fully account for thermal effects.

To facilitate a more direct comparison with the analytical models of \cite{Taylor2024}, we present an additional figure (Figure \ref{fig:col-density-3mj}) that focuses on the 3 Jupiter-mass case. This figure compares our simulation results for different disk aspect ratios (H/R = 0.05, 0.075, and 0.1) with the theoretical models from \cite{Taylor2024}. We normalized the density so that our curves also go through the common point of the analytic model. For this comparison, we calculated the column density starting outside the circumplanetary region, effectively removing the pressure-dominated part to align with the assumptions in the \cite{Taylor2024} models. In Figure \ref{fig:col-density-3mj}, we observe that our pressure-free profiles show a consistent pattern across different disk aspect ratios, with a relatively uniform column density across most polar angles in the log scale and a sharp increase towards the disk midplane. This pattern most closely resembles the ``equatorial'' case from \cite{Taylor2024} rather than their polar infall model. The discrepancy in shapes could be attributed to differences in the assumed inflow rates or the effects of long-term evolution in our simulations that may lead to more mass accumulation in the CPD.

\end{document}